\documentclass[twocolumn]{aastex631}

\usepackage{mathtools,amsmath} 
\usepackage{graphicx}
\usepackage{xcolor}
\usepackage{txfonts}
\usepackage{comment}
\usepackage{booktabs}  
\usepackage{csvsimple}
\usepackage[]{hyperref}
\def\teff{{\rm\,T_{eff}}}

\begin{document}

\title{Full abundance study of two newly discovered barium giants}

\author[0000-0001-5412-869X]{Sara Vitali}
\affiliation{Instituto de Estudios Astrofísicos, Facultad de Ingeniería y Ciencias, Universidad Diego Portales, Av. Ejército Libertador 441, Santiago, Chile}
\affiliation{Millenium Nucleus ERIS }
\email{sara.vitali@mail.udp.cl}

\author[0000-0003-3833-2513]{Ana Escorza}
\affiliation{Instituto de Astrofísica de Canarias, C. Vía Láctea, s/n, 38205 La Laguna, Santa Cruz de Tenerife, Spain}
\affiliation{Universidad de La Laguna, Dpto. Astrofísica, Av. Astrofísico Francisco Sánchez, 38206 La Laguna, Santa Cruz de Tenerife, Spain.}

\author[0000-0003-4538-9518]{Ditte Slumstrup}
\affiliation{European Southern Observatory, Alonso de Córdova 3107, Vitacura, Santiago, Chile}

\author{Paula Jofré}
\affiliation{Instituto de Estudios Astrofísicos, Facultad de Ingeniería y Ciencias, Universidad Diego Portales, Av. Ejército Libertador 441, Santiago, Chile}
\affiliation{Millenium Nucleus ERIS }

\begin{abstract}

Barium (Ba) stars are chemically peculiar stars that show enhanced surface abundances of heavy elements produced by the slow-neutron-capture process, the so-called s-process. These stars are not sufficiently evolved to undergo the s-process in their interiors, so they are considered products of binary interactions. Ba stars form when a former Asymptotic Giant Branch (AGB) companion, which is now a white dwarf, pollutes them with s-process-rich material through mass transfer. This paper presents a detailed chemical characterization of two newly discovered Ba giants. Our main goal is to confirm their status as extrinsic s-process stars and explore potential binarity and white dwarf companions. We obtained high-resolution spectra with UVES on the Very Large Telescope to determine the chemical properties of the targets. We perform line-by-line analyses and measure 22 elements with an internal precision up to 0.04 dex. The binary nature of the targets is investigated through radial velocity variability and spectral energy distribution fitting. We found that both targets are enhanced in all the measured s-process elements, classifying our targets as Ba giants. This is the first time they are classified as such in the literature. Additionally, both stars present a mild enhancement in Eu, but less than in pure s-process elements, suggesting that the sources that polluted them were pure s-process sources. Finally, we confirmed that the two targets are RV variable and likely binary systems. The abundances in these two newly discovered polluted binaries align with classical Ba giants, providing observational constraints to better understand the s-process in AGB stars.

\end{abstract}


\section{Introduction}\label{sect:intro}

Mass transfer in low- and intermediate-mass binary systems is known to lead to the formation of several families of chemically peculiar stars. For example, barium (Ba) stars \citep{BidelmanKeenan51}, CH stars \citep{Keenan42}, CEMP-s stars (Carbon-Enhanced Metal-Poor stars enhanced in heavy metals, \citealt{BeersChristlieb05}) and Tc-poor S stars \citep{SmithLambert90} show in their atmospheres peculiar overabundances of elements synthesised by the slow neutron-capture process (s-process; \citealt{Burbidge57, Kappeler11,2023Lugarorew}). The s-process occurs at the end of the Asymptotic Giant Branch (AGB) phase, during the thermal pulsing phase (TP-AGB), where the thermal pulses mix material from the core into the envelope \citep[e.g.][]{Lugaro03long, Cristallo09, Karakas10, KarakasLattanzio14}. The aforementioned chemically peculiar stellar families are not luminous enough to have synthesized these s-process elements themselves. However, through years of radial-velocity monitoring programs \citep[e.g.][]{McClure84, Udry98, Jorissen98, North00, Jorissen19, Escorza19}, it is well-established that their chemical peculiarities are caused by mass transfer from a former AGB companion with an envelope rich in heavy elements after going through the TP-AGB. The companion evolved off the AGB phase long ago and is now a faint and cool white dwarf (WD). 

In addition to the indirect evidence provided by the surface chemistry \citep[e.g.][]{Lugaro03, Lugaro12, Lugaro16, Cseh18, Karinkuzhi18}, a few Ba star systems show excess UV flux attributable to a WD \citep{Bohm-Vitense84, Bohm-Vitense00, Gray11}. In most systems, the companions are cool and not directly detectable, but their masses, derived combining radial velocity and astrometric data, are consistent with that of white dwarfs \citep[e.g.][]{Pourbaix2000, EscorzaDeRosa23}. Additionally, the ranges covered by the orbital parameters of Ba and related stars are well-determined \citep[e.g.][and references therein]{Escorza19, Jorissen19}. However, the exact mass-transfer mechanisms involved in their formation are not well understood \citep[e.g.][]{ToutEggleton88, Pols03, Mohamed07, Izzard10, Abate18, Saladino19, Gao22}.

Detailed chemical studies of s-process polluted stars can be helpful to learn about the nucleosynthesis processes that took place in their former AGB companions \citep[e.g.][]{Husti09, Shejeelammal20, Cseh22, denHartogh23} and to constrain parameters related to the mass transfer episode between the AGB star and the current polluted object \citep[e.g.][]{Stancliffe21}. In this paper, we present a detailed spectroscopic analysis of two Ba giants that had not been identified as such in the past. Section \ref{sect:target} includes information about the two targets and about the spectra we used. In Sect. \ref{sect:methods}, we described our methodology and provided information about the spectral lines used for the determination of the heavy metal abundances. In Sect. \ref{sect:results}, we present our results, and discuss them in the context of Ba giants. There we also investigate binarity in the two targets, since they are expected to be binaries by formation, but have not been flagged as such either. Finally, in Sect. \ref{sect:conclusions}, we summarise our findings.

\section{Target selection and spectroscopic data}\label{sect:target}
\subsection{Target selection}
This paper focuses on two chemically peculiar stars: 2MASS\,J04034842+1551272 (also HD\,285405) and 2MASS\,J16564223-2108420. The targets were part of the initial sample of \citet{2024Vitali} (hereafter V24) since they were thought to have followed standard single-star evolution. However, the analysis revealed that they were anomalously enhanced in several heavy metals, and they were thus discarded for the purpose of that study. Since the initial target selection, newer versions of the APOGEE and GALAH catalogues have been released, which also show signs of heavier element enhancement. This confirmed the initial indication of anomalous abundance patterns and motivated the full abundance study based on high-resolution high-signal-to-noise spectra that we present here. 

The first parts of Table \ref{tab:AP} list several relevant identifiers for the two stars, their coordinates, their Gaia G magnitude \citep{GaiaEDR3summary} and their distances above the Galactic plane. As expected for Ba stars, they are both classic disc members \citep{Mennessier97}. The two stars were targeted in the \textsl{K2} mission \citep{2014PASP..126..398H}, as the rest of the initial sample studied by V24. The third part of the table lists their asteroseismic parameters $\nu_{\rm max}$ (frequency of maximum power), $\Delta \nu$ (large frequency separation) from the SYD pipeline \citep{Huber2009}, the resulting surface gravity ($\log g$; see Sect.~\ref{sect:methods} for more details), and the mass and radius obtained from seismic scaling relations (see Sect.~\ref{sect:methods}). As mentioned above, the targets are also part of recent data releases of the APOGEE \citep{APOGEEsurvey} and GALAH \citep{GALAHsurvey} spectroscopic surveys. We used these catalogues to investigate the radial velocity (RV) variability of the targets in Sect. \ref{sect:Bastars}.

\subsection{UVES spectroscopic data}

The spectra were obtained under the ESO programme ID 108.22DX with the UVES cross-dispersed echelle spectrograph \citep{UVES}, mounted the Unit Telescope 2 of the Very Large Telescope (VLT), at the Paranal Observatory. The resolving power of the setup employed is $R \sim 110\,000$, with wavelength range 480-680\,nm. The signal-to-noise ratio (SNR) of the two spectra are 119 for J04034842+1551272 and 80 for J16564223-2108420. The spectra were reduced with the ESO UVES pipeline \citep{UVESpipeline, 2013Freudling}.

\section{Spectral analysis}\label{sect:methods}

The atmospheric parameters and individual abundances were derived using the public spectral software \texttt{iSpec} \citep{2014Blanco, 2019Blanco}. We followed the methodology outlined in \cite{2019Blanco,2020Casamiquela} and V24. We summarise our steps in the next sections, but we refer to the listed papers for additional details.

\subsection{Stellar parameters}\label{sect:atm_param}

We used \texttt{iSpec} to fit our observed spectra with synthetic spectra generated on the fly. We chose the radiative transfer code Turbospectrum \citep{1998A&A...330.1109A,2012ascl.soft05004P}, which considers local thermodynamic equilibrium and the one-dimensional spherical MARCS model atmospheres \citep{2008A&A...486..951G}. Additionally, we relied on the line list of the \textit{Gaia}-ESO Survey (GES, \citealt{2021Heiter}). For the atmospheric parameters, we followed the line selection illustrated in the work of \cite{2019Blanco}, which avoids features such as blends, telluric contamination, or continuum displacement during the line selection process.

Adopting the same strategy as in V24, we determined the effective temperature, surface gravity, metallicity, and broadening parameters for our two targets as a first step. Then, as done in V24, we used that effective temperature to calculate a seismic mass, radius and $\log\,g$ relying on the seismic scaling relation from \cite{1991Brown} and \cite{1995Kjeldsen}. That seismic $\log\,g$ value was subsequently fixed throughout the spectral analysis. The differences between the initial spectroscopic and seismic $\log\,g$ values are 0.04 dex for J04034842+1551272 and 0.19 dex for J16564223-2108420. The larger discrepancy for the latter star may be attributed to its colder effective temperature and lower metallicity, however, such differences do not affect the final line-by-line measured abundances strongly. We chose to use the asteroseismic $\log\,g$ to ensure consistency with the analysis performed in V24. Additionally, various studies have demonstrated that asteroseismology provides one of the most accurate methods for determining the surface gravity parameter, with an accuracy of 0.02–0.05 dex \citep{2013Creevey,2016Brogaard}.

\begin{figure}[t]
\centering
  \includegraphics[width=1\columnwidth]{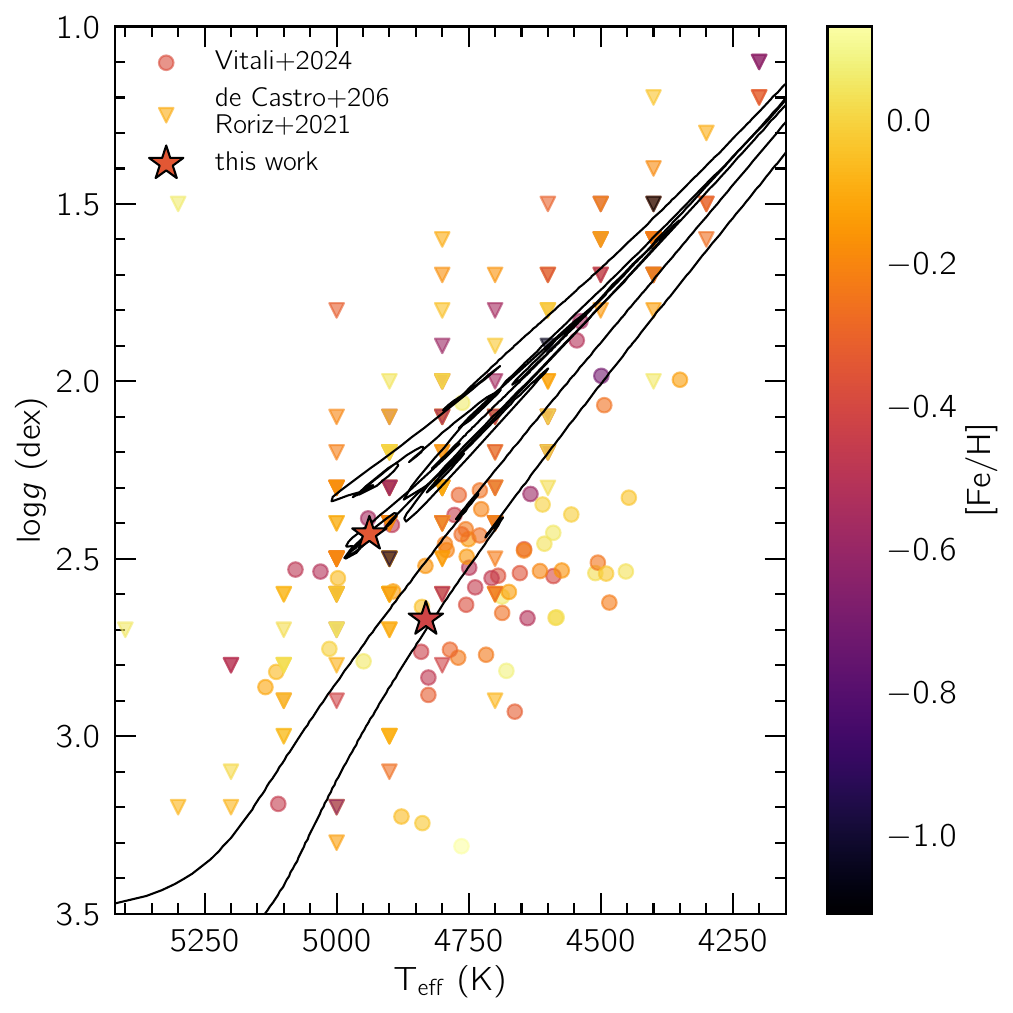} 
 \caption{Effective temperature — surface gravity diagram (Kiel diagram) including our target stars J04034842+1551272 and J16564223-2108420. We overplot them with the sample of effectively single red giants studied by V24 (circles), and the sample of Ba giants from \cite{deCastro16, 2021RorizRb, Roriz21} (triangles). The three samples are colour-coded as a function of each target's metallicity. We also overplotted evolutionary tracks computed with the STAREVOL code \citep{Siess00, Siess08} for [Fe/H]~=~$-0.5$ and initial masses equal to 1.0 and 1.5~M$_{\rm \odot}$. }
 \label{fig:iso}
\end{figure}

The best-fit results are presented in Table \ref{tab:AP}. Figure \ref{fig:iso} shows the targets on a Kiel diagram using the seismic $\log\,g$ and the spectroscopic effective temperature.

\renewcommand{\arraystretch}{0.95}
\setlength{\tabcolsep}{2pt}
\begin{table}[]
\caption{Identifiers, coordinates, G magnitudes, asteroseismic quantities, stellar photospheric parameters and chemical abundances, and radial velocities derived for the target stars. The surface gravity was computed using asteroseismic scaling relations and kept fixed during the spectral analysis, hence, it is not listed among the spectroscopic parameters. The masses and photospheric radii listed are also derived from seismic scaling relations.}\label{tab:AP}
    \centering
    \begin{tabular}{l c c} 
    \hline \hline
        Parameter & J04034842+1551272 & J16564223-2108420\\
    \hline \hline
        APOGEE ID & 2M04034842+1551272 & 2M16564223-2108420\\
        GALAH ID & 140824006301237 & 170506004901324\\
    \hline\hline
        \multicolumn{3}{c}{Coordinates and Magnitudes}\\
    \hline   
        RA & 04 03 48.42 & 16 56 42.240\\
        Decl. & +15 51 27.20 & -21 08 42.05\\
        Gmag & 10.789$\pm$0.003 & 11.376$\pm$0.003\\
        Galactic z [kpc]  & -0.376 & 0.196 \\
    \hline\hline
        \multicolumn{3}{c}{Asteroseismic parameters}\\
    \hline
        $\nu_{\rm max}$ & 41.34 $\pm 3.25 $ & 58.11$\pm 4.79 $\\
        $\Delta \nu$ & 3.96 $\pm 0.15$ & 6.29$\pm 0.05$ \\
        $\log\,g~\mathrm{[dex]}$ & 2.43$\pm 0.02$ & 2.67$\pm 0.02$\\
        $\rm{M [M_{\odot}]}$ & 1.1 $\pm 0.1$ & 1.0$\pm 0.05$ \\     
        $\rm{R [R_{\odot}]}$ & 10.3 $\pm 0.2$ & 7.6$\pm 0.1$ \\ 
    \hline\hline
        \multicolumn{3}{c}{Spectroscopic parameters}\\
    \hline        
        $\teff~\mathrm{[K]}$ & 4938$\pm 15$ & 4831$\pm 13$\\
        $[\mathrm{Fe/H}]~\mathrm{[dex]}$ & $-0.34 \pm 0.01$ & $-0.42 \pm 0.01$\\
        $v_{\rm mic} \,[\mathrm{km s}^{-1}]$ & 1.42$\pm 0.02$ & 1.25$\pm 0.03$\\
        $v_{\rm mac} \,[\mathrm{km s}^{-1}]$ & 4.48$\pm 0.03$ & 4.33$\pm 0.03$ \\
    \hline\hline
        \multicolumn{3}{c}{Individual Abundances}\\
    \hline
        $[\mathrm{Na1/Fe}]~\mathrm{[dex]}$ & $ +0.19 \pm  0.02 $ & $ +0.11 \pm 0.02 $\\
        $[\mathrm{Mg1/Fe}]~\mathrm{[dex]}$ & $ +0.14 \pm  0.02 $ & $ +0.06 \pm 0.02 $\\
        $[\mathrm{Al1/Fe}]~\mathrm{[dex]}$ & $ +0.18 \pm  0.01 $ & $ +0.25 \pm 0.01 $\\
        $[\mathrm{Si1/Fe}]~\mathrm{[dex]}$ & $ +0.10 \pm  0.01 $ & $ +0.14 \pm 0.02 $\\
        $[\mathrm{Ca1/Fe}]~\mathrm{[dex]}$ & $ +0.11 \pm  0.01 $ & $ +0.12 \pm 0.01 $\\
        $[\mathrm{Sc2/Fe}]~\mathrm{[dex]}$ & $ +0.13 \pm  0.01 $ & $ +0.12 \pm 0.01 $\\
        $[\mathrm{Ti1/Fe}]~\mathrm{[dex]}$ & $ +0.05 \pm  0.01 $ & $ +0.13 \pm 0.02 $\\
        $[\mathrm{Ti2/Fe}]~\mathrm{[dex]}$ & $ +0.06 \pm  0.01 $ & $ +0.10 \pm 0.02 $\\
        $[\mathrm{V1/Fe}]~\mathrm{[dex]}$  & $ -0.01 \pm  0.02 $ & $ -0.05 \pm 0.01 $\\
        $[\mathrm{Cr1/Fe}]~\mathrm{[dex]}$ & $ +0.02 \pm  0.01 $ & $ +0.01 \pm 0.02 $\\
        $[\mathrm{Mn1/Fe}]~\mathrm{[dex]}$ & $ -0.04 \pm  0.02 $ & $ -0.07 \pm 0.01 $\\
        $[\mathrm{Co1/Fe}]~\mathrm{[dex]}$ & $ +0.01 \pm  0.01 $ & $ +0.00 \pm 0.01 $\\
        $[\mathrm{Ni1/Fe}]~\mathrm{[dex]}$ & $ -0.02 \pm  0.01 $ & $ -0.01 \pm 0.02 $\\
        $[\mathrm{Cu1/Fe}]~\mathrm{[dex]}$ & $ +0.09 \pm  0.01 $ & $ +0.08 \pm 0.02 $\\
        $[\mathrm{Zn1/Fe}]~\mathrm{[dex]}$ & $ +0.05 \pm  0.03 $ & $ +0.01 \pm 0.01 $\\
        $[\mathrm{Y2/Fe}]~\mathrm{[dex]}$  & $ +1.05 \pm  0.03 $ & $ +0.94 \pm 0.03 $\\
        $[\mathrm{Zr1/Fe}]~\mathrm{[dex]}$ & $ +0.74 \pm  0.02 $ & $ +0.64 \pm 0.01 $\\
        $[\mathrm{Ba2/Fe}]~\mathrm{[dex]}$ & $ +1.11 \pm  0.01 $ & $ +1.58 \pm 0.02 $\\
        $[\mathrm{La2/Fe}]~\mathrm{[dex]}$ & $ +1.05 \pm  0.02 $ & $ +1.18 \pm 0.03 $\\
        $[\mathrm{Ce2/Fe}]~\mathrm{[dex]}$ & $ +1.16 \pm  0.02 $ & $ +1.37 \pm 0.02 $\\
        $[\mathrm{Pr2/Fe}]~\mathrm{[dex]}$ & $ +1.16 \pm  0.03 $ & $ +1.47 \pm 0.04 $\\
        $[\mathrm{Nd2/Fe}]~\mathrm{[dex]}$ & $ +1.15 \pm  0.02 $ & $ +1.39 \pm 0.01 $\\
        $[\mathrm{Eu2/Fe}]~\mathrm{[dex]}$ & $ +0.32 \pm  0.02 $ & $ +0.46 \pm 0.02 $\\
    \hline\hline
        \multicolumn{3}{c}{Radial velocity information}\\
    \hline
        BJD$_{\rm obs}$ & 2459532.128 & 2459478.042\\
        RV~$[\mathrm{km s}^{-1}]$ & 22.25$\pm$0.15 & -20.86 $\pm$0.13\\
    \hline\hline
    \end{tabular}
\end{table}

\subsection{Chemical abundances}\label{sect:chemistry}
Adopting the same method described in V24 we derived the individual abundances of $\alpha-$elements (Mg, Si, Ca, Ti), odd-Z elements (Na, Al, Sc, V, Cu), iron-peak elements (Cr, Mn, Fe, Co, Ni, Zn), and neutron-capture elements (Y, Zr, Ba, La, Ce, Pr, Nd, Eu). The abundances of key light elements in Ba stars, such as carbon, nitrogen, and oxygen, could not be determined due to the absence of characteristic features within our wavelength range or the insufficient quality of the fits obtained. The final mean abundance ratios of the line-by-line absolute abundances of each element are listed in Table \ref{tab:AP}. The reference solar abundances were derived as described in V24 by computing line-by-line differential abundances from a solar spectrum of similar resolution, obtained from the library of \cite{Blanco-Cuaresma2014}. The uncertainties reported are computed by determining the average dispersion of our abundance measurements after perturbing ten times the spectra within their flux errors. This evaluates the internal precision of our analysis, which is sufficient for the purpose of classifying our stars as Ba giants.

For most elements, we relied on the same line selection as in V24. However, since this study focuses on heavy elements, we revisited the line selection for neutron-capture elements to improve their abundances. Additionally, this more careful treatment of the neutron-capture elements allowed us to expand the number of elements studied with respect to those derived by V24. A complete table providing the line selection is presented in Appendix \ref{sec:append} (Table \ref{tab:linelist}). 

The entire analysis was done in 1D local thermodynamic equilibrium (LTE), taking the hyperfine structure (HFS) into account. The adopted HFS components and isotopic splitting values are provided in the line list of \cite{2021Heiter}. We note that while there are non-LTE (NLTE) corrections for s-process elements in the literature (e.g. \citealt{2012Bergemann, Karinkuzhi18, 2021Cescutti}), these are tailored to normal, non-enhanced stars. To our knowledge, NLTE corrections have not been derived for highly s-process-enhanced stars like the two presented here. In any case, the absorption lines we used (except for barium lines) are all in the linear regime. Therefore, we can rely on the corrections derived for non-enhanced stars. For our targets' atmospheric parameter and metallicity ranges, NLTE corrections are on the order of 0.1\,dex or less \citep{2012Hansen, 2017Mashonkina, 2023Alexeeva}. Given the strong s-process element enhancements we derived ($\gtrsim$1\,dex, Table \ref{tab:AP}), these corrections will not change the nature of our targets as Ba stars. Thus, to maintain consistency when comparing our results with other LTE results reported in the literature, we provide and discuss LTE chemical abundances only.

\subsubsection{Light s-process elements: Y, Zr}\label{sect:light-s}

The yttrium (Y) abundance measurements for the programme stars were derived using the Y II lines listed in Table \ref{tab:linelist}. As discussed above, NLTE corrections are available for this element in the case of giant stars, but are negligible ($<$ 0.05 dex) for our metallicity range \citep{2023Alexeeva, 2023Storm}.

Our measurement for zirconium (Zr) is based on five different Zr I lines available in our spectra. For the analysis done in V24, we derived the Zr abundance using a single Zr II line at 535.009 nm, which proved challenging to model. To ensure a consistent comparison with this study and with the work of \cite{deCastro16}, who based their measurements on Zr I lines, we computed new Zr I abundances for the entire UVES sample from V24 and updated the online catalogue.

Although strontium (Sr) is a typical light s-process element, we are not able to provide the Sr surface abundance of our targets in our final results. Based on the work done on Ba stars by \cite{Karinkuzhi18, Karinkuzhi21}, \cite{2023Roriz}, and others, the most commonly used Sr lines fall outside the wavelength range of our spectra. A visual inspection of our fits confirmed that the lines available at wavelengths redder than 480 nm are not useful for this analysis.

\subsubsection{Heavy s-process elements: Ba, La, Ce, Pr, Nd}\label{sect:heavy-s}

\begin{figure}[t]
\centering
 \includegraphics[width=\columnwidth]{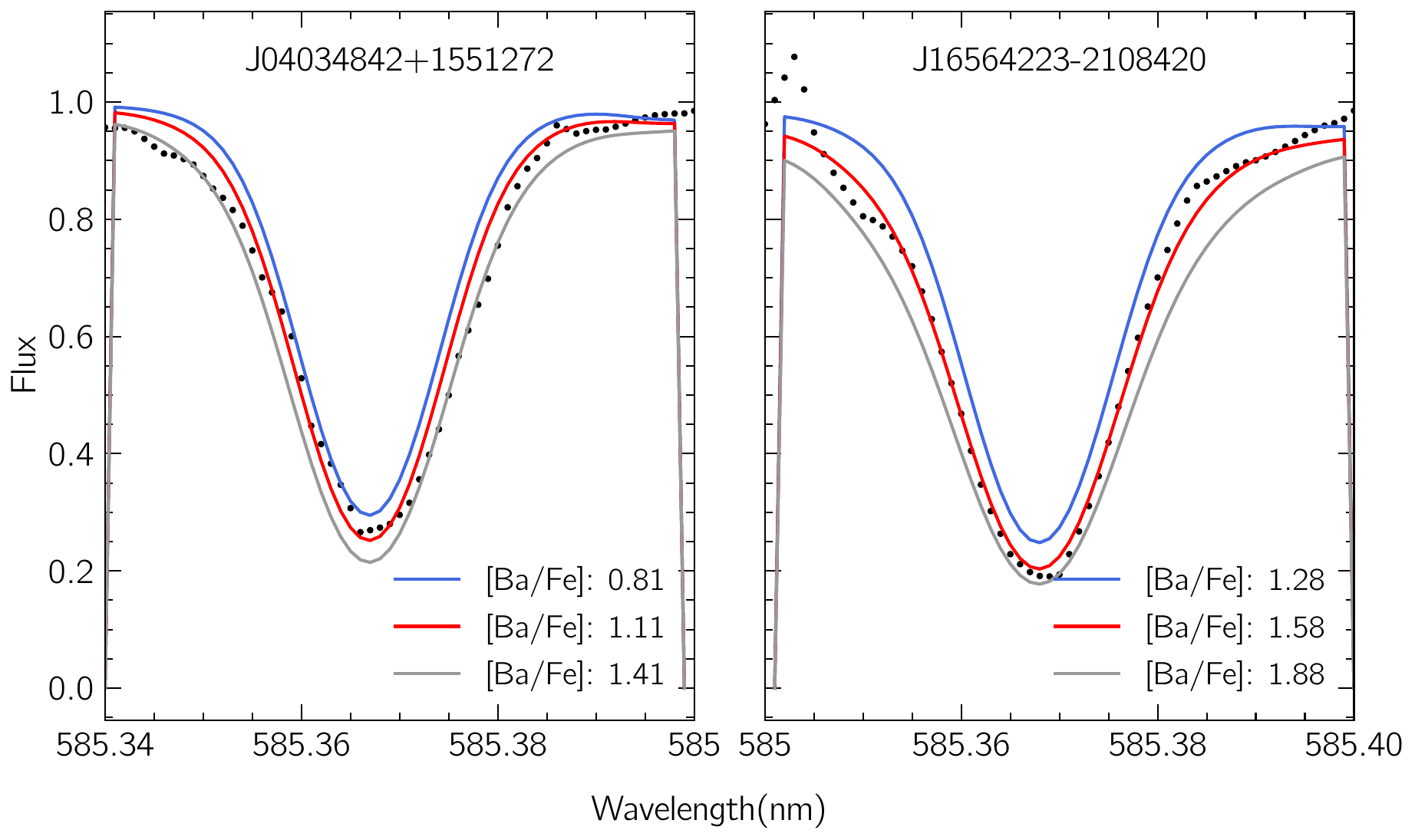} 
 \caption{Synthesis of the Ba spectral line at 585.366 nm compared with the observed line (black dots) for both targets. The red line represents a synthetic model created with our final parameters and [Ba/Fe] abundance (see Table \ref{tab:AP}). The blue and pink lines show synthetic spectra with [Ba/Fe] varied by $\pm$0.3 dex, while the grey spectrum is synthesised with the Ba solar abundance value. This comparison highlights the strength of the Ba line. Even though its enhancement makes the variation less obvious, the sensitivity of the line profile is visible, especially in the wings. This behaviour demonstrates the complexity of this measurement.}
 \label{fig:ba_lines}
\end{figure}

It is known that for giant stars the Ba II lines can be very strong and affected by saturation and hyperfine splitting \citep{2003Mashonkina, 2020Liu}. 
The isotopic splittings and HFS constants used in this analysis are taken from the GES line-list, which references the measured values reported by  \cite{1984Wendt} and \cite{1986Silverans}. At the same time, works such as those by \cite{2015Korotin, 2009Andrievsky} and \cite{2019Eitner} demonstrated that at our metallicity range, the deviations from LTE of the Ba II line at 585.36 nm are minor, at most up to 0.1~dex. With this in mind, we derived Ba II abundances using only the mentioned line. However, we noted that our best-fitting models do not fully reproduce the observed line profile. Therefore, to validate the [Ba/Fe] results, we created three additional synthetic spectra: one with [Ba/Fe]~=~0 and two more by varying the Ba abundance by $\pm$0.3\,dex, and they are displayed in Fig.~\ref{fig:ba_lines}. The red line represents the spectrum synthesized using our final [Ba/Fe] results, and the grey spectrum adopts the solar abundance value for Ba. Due to the high abundance ratios derived for our targets, we found that a variation of up to 0.3 dex is necessary to show a noticeable change in this Ba line. As discussed above, the reported uncertainties still correspond to the internal precision only, but a variation up to 0.3~dex is possible. This, however, would not alter the chemically peculiar nature of the programme stars. It is also important to emphasise that the accuracy of barium abundances is generally affected by measurement difficulties, primarily because the strength of these lines places them outside the linear regime, as clearly depicted in Fig.~\ref{fig:ba_lines}

Lanthanum (La) lines are known to show significant hyperfine splitting. This was taken into account following the prescription by \cite{Lawler2001}. Five La II lines were measured covering most of our spectral range similar to cerium (Ce), for which measurements were obtained using six Ce II lines. Praseodymium (Pr) abundances were derived from two well-fitted Pr II lines. For neodymium (Nd), we measured four Nd II lines, mostly located in a confined spectral window from 529 to 531 nm. For all the above-mentioned elements (La, Ce, Pr and Nd) the spectral lines are only known to be minimally affected by NLTE effects \cite[see e.g.,][]{Maiorca2011,2017Abdelkawy,2020Shaltout}, never exceeding 0.1\,dex. All these lines are listed in Table \ref{tab:linelist}.

Among s-process elements, niobium (Nb) and zirconium (Zr) have been identified as effective proxies for the operation temperatures of the s-process in the former AGB companions \citep{2015Neyskens}. The authors proposed using the [Nb/Fe] vs. [Zr/Fe] ratios to distinguish stars with an extrinsic s-process enhancement from intrinsically enhanced AGB stars. Subsequent studies, such as those by \cite{Karinkuzhi18, Shetye18, Roriz21}, validated this method as a robust tool for characterising the nature of s-process-rich stars. Although there are Nb lines present in our covered wavelength range, they are too weak to derive the Nb abundance and/or strongly affected by blends. This, unfortunately, leaves us unable to use the [Nb/Fe] vs. [Zr/Fe] characteristic to distinguish between an extrinsic or an intrinsic enhancement. Nonetheless, the Kiel diagram in Fig. \ref{fig:iso} clearly indicates that our targets are in the red giant phase and have not yet evolved onto the AGB, which means they cannot yet be intrinsically enhanced. Hence, their s-process enhancement must originate from a companion that previously synthesised these elements during its AGB phase.

\vspace{0.355cm}
\begin{figure}[t]
\centering
 \includegraphics[width=\columnwidth]{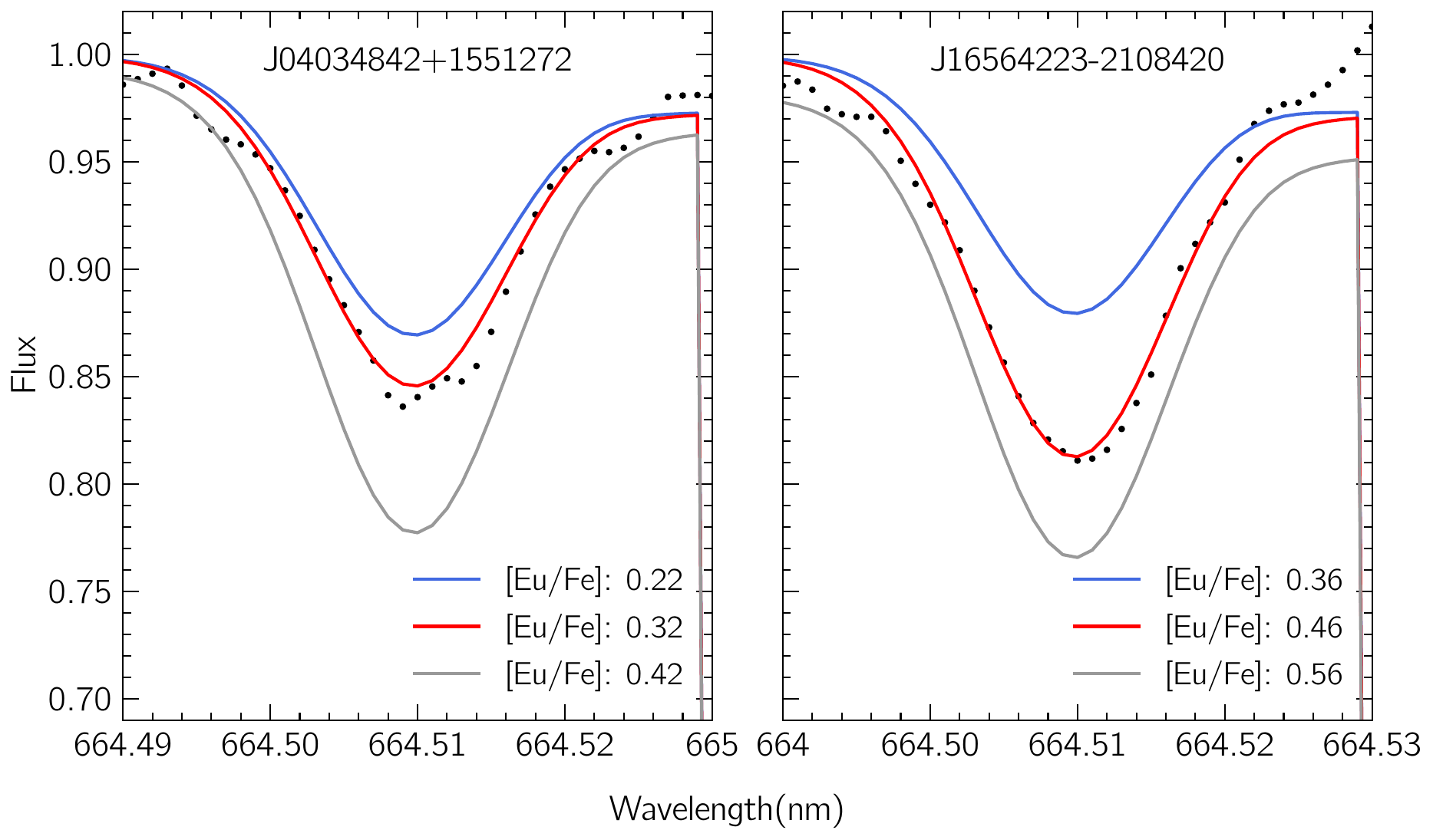} 
 \caption{Similar plot to Fig.~\ref{fig:ba_lines} but for the Eu II line at 664.51 nm. Synthetic and observed data follow the same colours as for the Ba II line. This time the variation among the three Eu-enhanced spectra corresponds to 0.1 dex. Again, the fourth spectrum is synthesised with the Eu solar value. The red line refers again to the [Eu/Fe] result presented in Table \ref{tab:AP}. The sensitivity of the line to abundance changes highlights the accuracy of our spectral fit, even with the blend impacting this Eu II line. }
 \label{fig:eu_lines}
\end{figure}

\subsubsection{r-process elements: Eu}

For the family of r-process elements, our results are based on a single line of ionized europium (Eu). Specifically, the Eu II line used in the analysis is located in the redder part of the spectrum at 664.51 nm (Fig.~\ref{fig:eu_lines}). The majority of the lines typically used to derive Eu abundances with optical spectra are in the bluer part ($\lessapprox420$ nm). For the line at 664.51 nm, in the case of stars in our metallicity range, the NLTE corrections are very small \citep{2014Mashonkina,2024Storm}. 
The inclusion of hyperfine splitting \citep{1992Villemoes} and isotopic corrections \citep{Lawler2001} reduces the impact of blending effects, which can make the determination of a robust Eu abundance challenging. Given these circumstances, we repeated the exercise performed for the Ba lines in Fig. \ref{fig:ba_lines}. Figure \ref{fig:eu_lines} illustrates that a deviation of $\pm$0.1 dex is sufficient to reveal a clear divergence from the best-fit model. Consequently, the [Eu/Fe] ratios are validated despite the impact of hyperfine splitting and blending effects on Eu II lines, the latter likely caused by a Cr I line on the red wing. 

Attempts were made to increase the number of r-process elements measured in the program stars. Unfortunately, the samarium (Sm) lines present in our spectral range did not provide reliable measurements due to numerous blends. Concerning other r-process elements such as Dy, Gd, Er, Hf, Os, and Ir, all their lines are around or below 420 nm, hence, outside our wavelength coverage.

\begin{figure*}[t]
\centering
 \includegraphics[width=\textwidth]{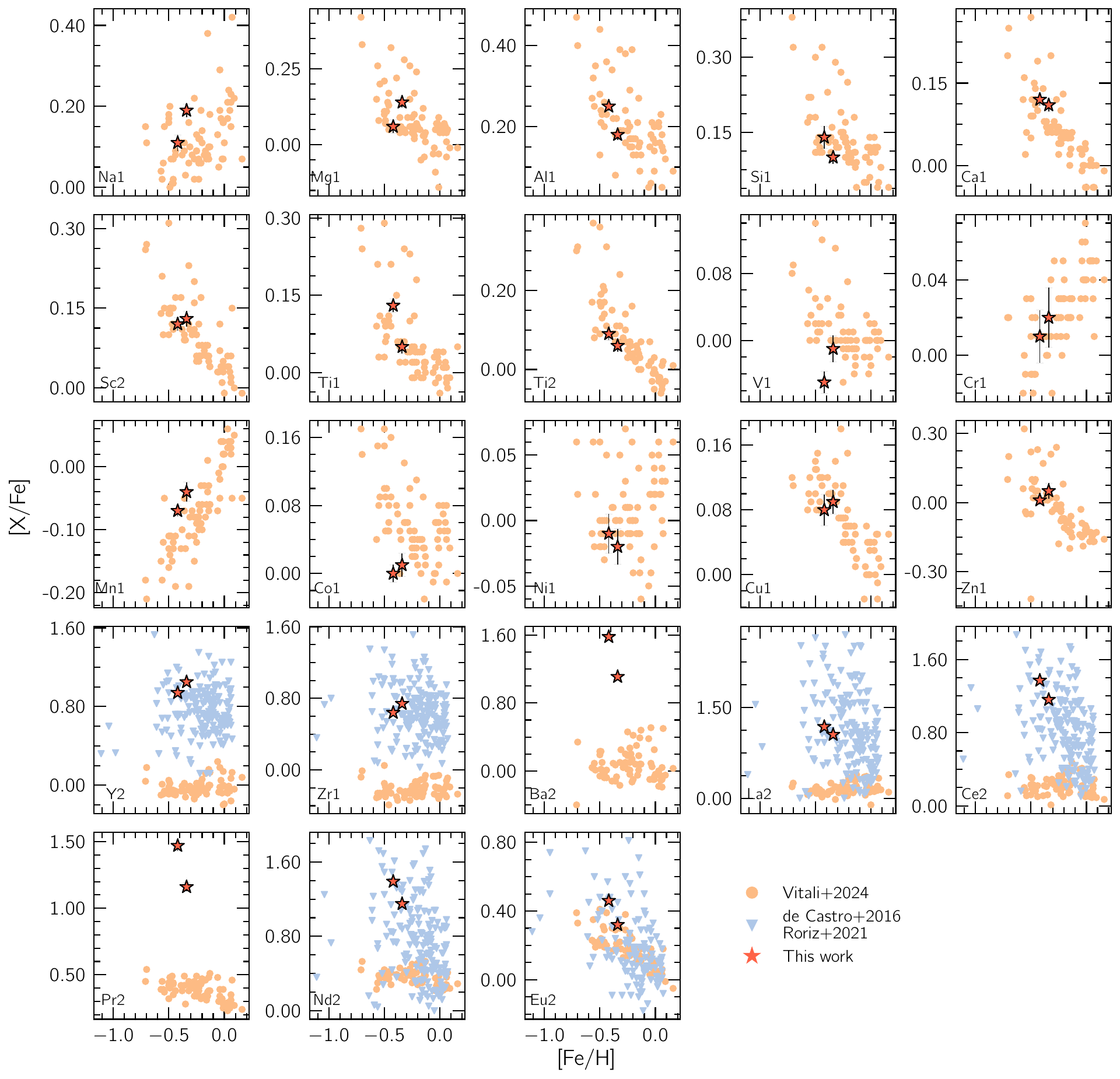} 
 \caption{[X/Fe] as a function of metallicity for our targets (red stars). For comparison, we also include the sample of effectively single red giants from V24 (orange circles) and the Ba giants from \cite{deCastro16}, with Eu values added from \cite{Roriz21} (light blue triangles)}
 \label{fig:el_feh}
\end{figure*}

\begin{figure}[t]
\centering
 \includegraphics[width=0.5\textwidth]{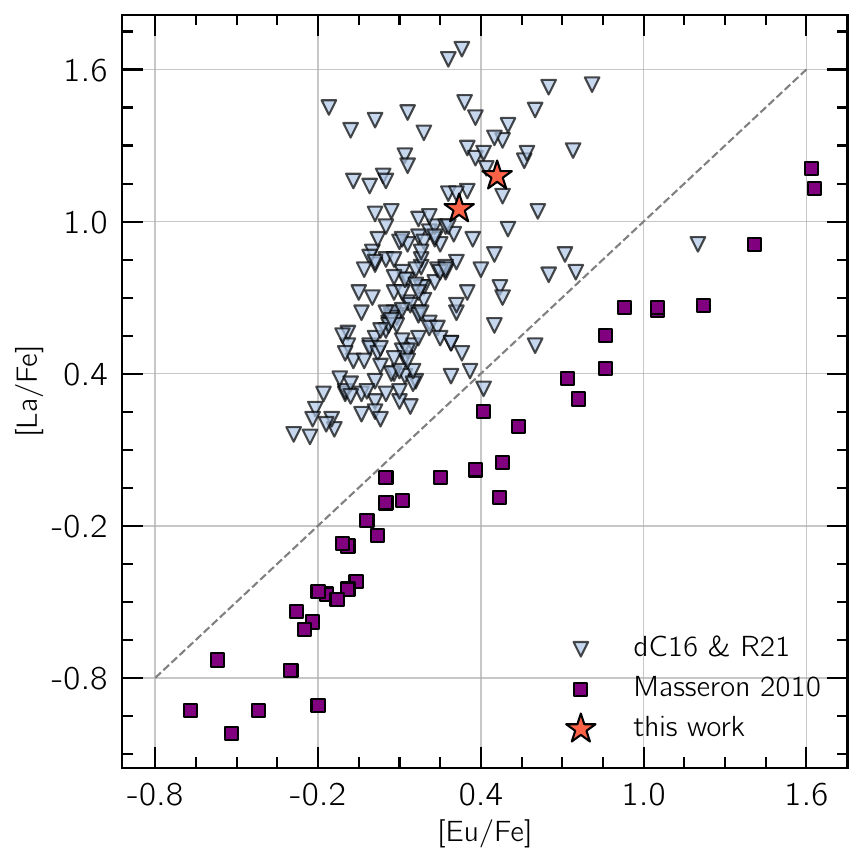} 
 \caption{[La/Fe] as a function of [Eu/Fe] for our target stars to confirm their pure-s-process pollution. For comparison, the light blue triangles represent the sample of well-known Ba giants from \cite{deCastro16, Roriz21} and the purple squares show a sample of r-process rich CEMP stars studied by \cite{Masseron2010}.}
 \label{fig:eu_la}
\end{figure}

\section{Results and discussion}\label{sect:results}

\subsection{Stellar parameters and abundances}

As shown in the two last rows in Fig.~\ref{fig:el_feh}, the two target stars are clearly enhanced in s-process elements, with abundances up to more than 1 dex. In this section, we will compare our parameters and abundances with two reference samples. The first one is the sample of effectively single pulsating red giants studied by V24 (orange circles), and the second one is a sample of well-known Ba giants published by \cite{deCastro16} and complemented in the last few years by \cite{2021RorizRb, Roriz21} and \cite{2024RorizTungsten}, who added measurements for europium (Eu). Both the literature sample from \cite{deCastro16} and \cite{Roriz21} are represented by blue triangles. Figure \ref{fig:iso} shows our target stars in a Kiel diagram together with the two reference samples. All stars in the plot are colour-coded as a function of their metallicity. To guide the eye, we overplotted two evolutionary tracks for stars with initial masses equal to 1.0 and 1.5~M$_{\rm \odot}$ computed with the STAREVOL code \citep{Siess00, Siess08} for [Fe/H]~=~$-0.5$ (see \cite{Escorza17,Escorza19} for specific details about the grid). The masses of the program stars listed in Table \ref{tab:AP} are derived from scaling relations using our measured asteroseismic parameters. 

Fig. \ref{fig:iso} shows that our targets do not stand out considering effective temperature, surface gravity and metallicity. However, Fig.~\ref{fig:el_feh} shows the abundances derived for 22 chemical elements, where the anomalous abundance pattern stands out clearly. While the abundances of our targets are in line with those of normal giants for non-neutron-capture elements, J04034842+1551272 and J16564223-2108420 have significantly higher s-process abundances. We note that the sample from \cite{deCastro16} does not have Ba or Pr, but we included these panels as well to highlight the enhancement of our targets with respect to the normal red giants. As discussed in Sect. \ref{sect:heavy-s} and shown in Fig. \ref{fig:ba_lines}, Ba abundances are particularly difficult to measure. Nevertheless, we can still infer a final result that confirms the Ba enhancement in the two stars.

Figure \ref{fig:eu_la} shows our targets in the [La/Fe] versus [Eu/Fe] plane. This is a commonly used diagnostic to determine if the heavy metal abundances observed in a star have a slow- or rapid-neutron-capture origin. La is a representative element of the s-process, while Eu is representative of the r-process. When a star is enhanced in La and to the left of the diagonal line shown in the plot, one can assume that its heavy metals were mainly synthesized by the s-process. As expected, the Ba giants studied by \cite{deCastro16, 2021RorizRb, Roriz21, 2024RorizTungsten} mostly occupy this region. On the other hand, when a star is to the right of the diagonal, one can expect a significant contribution from the r-process. For reference, we added to the plot a sample of r-process-rich CEMP stars from \cite{Masseron2010}. These stars are classified as rI and rII in their work and are not expected to be part of binary systems. Even though our targets mildly enhanced in Eu, they are quite far from the diagonal and well-mixed with the sample of known Ba giants. This indicates that they were polluted by pure s-process-rich material and are very likely to belong to binary systems.

The high [Eu/Fe] ratios observed in our stars can be partly attributed to s-process sources. For instance, \cite{2006Allen} found that the r-process abundance levels are higher in Ba stars compared to "normal" stars due to the s-process formation chain, which includes the production of r-elements. However, it is well established that only approximately 5\% of Eu originates from s-process sources \citep{2015Bisterzo,2020Prantzos}. Other studies invoked the scenario where two types of pollution are responsible for the r-process level: pre-enrichment from a type-II supernova (SNII) followed by s-process enrichment from an AGB star \citep{2006Jonsell,2006MmBisterzo,2014Cui}. 
In our case, it is possible that our stars formed in molecular clouds that were initially polluted by SNII, responsible for the r-process enrichment, and then received mass from the AGB companion. However, one should also consider the intermediate-neutron-capture process (i-process), with intermediate neutron concentrations between the s- and r-processes, and first proposed by \cite{i-process}. This process used to be invoked only to explain the existence of CEMP-r/s stars \citep[e.g.][]{2006Jonsell, 2012Lugaro, Roederer16, Karinkuzhi21}, at lower metallicities than our targets, but it has recently been discovered to potentially operate at near-solar metallicity too \citep{Karinkuzhi23, Choplin24}.

To gain a deeper understanding of the s-process contributions and neutron exposures, one can use the heavy-to-light (\textit{[hs/ls]}) indexes of s-process elements. These chemical ratios are also valuable proxies for estimating the mass of the polluting AGB star. Different origins and formation pathways have been proposed for these two categories of elements. The \textit{hs} elements (Ba, La, Ce, Pr, and Nd) are known to primarily originate from low-mass AGB stars with masses $\leq 3 \mathrm{M_{\odot}}$ \citep{2003Lugaro, 2014Bisterzo}. In contrast, the \textit{ls} (Sr, Y, Zr) elements are mainly produced by more massive stars and intermediate-mass AGB stars \citep{Travaglio2004,2010ApJ...710.1557P,2023Goswami}. 

In this context we examined the \textit{[hs/ls]} for our programme stars. We computed the \textit{[hs/ls]} as the ratio of [\textit{hs}/Fe] to [\textit{ls}/Fe], where [\textit{hs}/Fe] and [\textit{ls}/Fe] are the averaged abundances of elements from the second s-process peak (Ba, La, Ce, Pr, Nd) and the first s-process peak (Y, Zr), respectively. The inferred ratios are 0.21 for J04034842+1551272 and 0.61 for J16564223-2108420. Utilising the FUll-Network Repository of Updated Isotopic Tables\& Yields (F.R.U.I.T.Y.) database \citep{2009Cristallo,Cristallo15} \footnote{\url{http://fruity.oa-teramo.inaf.it/}}, we found that model predictions for metallicities close to those of our targets (Z = 0.008 and Z = 0.006) suggest AGB star masses of 1.5 (for J04034842+1551272) to 3 (for J16564223-2108420) $\rm{M_{\odot}}$. Models with higher masses predict \textit{[hs/ls]} values that diverge more significantly from our measured ratios. Therefore, it is reasonable to conclude that the AGB companions were not more massive than 3 $\rm{M_{\odot}}$. This conclusion is further supported by the fact that our Mg measurements for these stars are in agreement with the values for normal giants following the Galactic local trend (see Fig. 4). Mg is expected to be enhanced in s-process stars polluted by more massive AGB stars when the $\prescript{22}{}{\mathbf{Ne(\alpha,n)}}^{25}\rm{Mg}$ reaction dominates over the $\prescript{13}{}{\mathbf{C(\alpha,n)}}^{16}\rm{O}$ reaction as the main provider of neutrons to feed the s-process \citep{2003Karakas,2012Longland,2014Fishlock}. Similar [Mg/Fe] values are reported for the eighteen Ba stars studied by \cite{Karinkuzhi18}, where the authors estimate the donor mass to be in the range of M$\sim\,2-3\,\rm{M_{\odot}}$.

\subsection{Binarity}\label{sect:Bastars}

Binarity is a prerequisite to form Ba and related s-process-polluted stars. As discussed in Sect. \ref{sect:intro}, when a star that is not luminous enough to be a TP-AGB star is found to be enhanced in s-process elements, one assumes it is polluted in a binary system and that it has a WD companion. However, sometimes this is difficult to confirm because the periods of these systems are long (see eccentricity-period diagrams in \citealt{Jorissen19, Escorza19, Escorza20}), and the WD companions are generally cool \citep[below 10\,000~K; e.g.][]{Bohm-Vitense00,Gray11}.

Our targets had not previously been identified as binaries. Their \textit{Gaia} DR3 Re-normalised Unit Weight Error (RUWE) values are 2.88 and 1.80 for J04034842+1551272 and J16564223-2108420, respectively \citep{GaiaEDR3summary}, which is not very high, but in both cases above the value of 1.4 often used to identify binary stars \citep[e.g.][]{Lindegren18, Kervella22}. Additionally, the non-single-star (NSS) \textit{Gaia} DR3 catalogue \citep{GaiaDR3binaries} did not include these targets. This is not surprising, however, since the data used in \textit{Gaia} DR3 covers about 1000 days \citep{GaiaDR3binaries} and the period distribution of Ba stars peaks at $\sim$2000~days, with some systems having 30\,000-day periods \citep[e.g.][]{Jorissen19, Escorza19, EscorzaDeRosa23}. Hence, to confirm the binarity of our targets, we studied the variability of their radial velocity data over a longer timespan than what Gaia offers at the moment (Sect. \ref{sect:RV}) and we searched for other pieces of evidence indicating the presence of a WD companion (Sect. \ref{sect:SED}).

\subsubsection{Radial velocity variability}\label{sect:RV}

\begin{figure}[t]
\centering
 \includegraphics[width=0.5\textwidth]{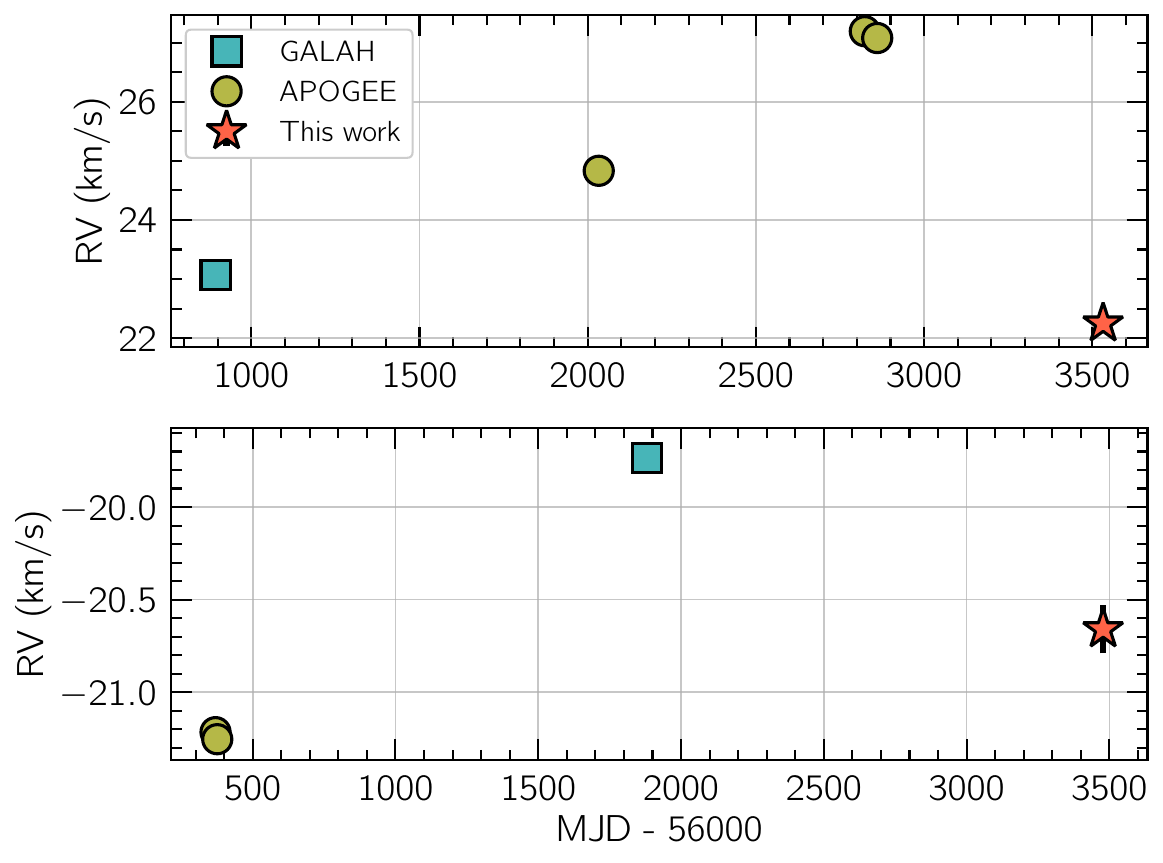} 
 \caption{RV data for J04034842+1551272 (top) and J16564223-2108420 (bottom) published by the GALAH \citep{GALAHsurvey} survey (blue square) and the APOGEE \citep{APOGEEsurvey} survey (olive circles) or measured for this work (red star). }
 \label{fig:RV}
\end{figure}

We only have one high-resolution UVES spectrum for each target. They were obtained in September and November 2021 for J16564223-2108420 and J04034842+1551272, respectively. The last section of Table \ref{tab:AP} lists the RV values that we derived from these spectra after applying the appropriate barycentric correction. The radial velocities were derived using a cross-match correlation algorithm \citep[e.g.][]{CCF} with the Arcturus line list designed by \cite{Arcturus} as template. To do this, we used \texttt{iSpec} as described by \cite{Blanco-Cuaresma2014}. 

In Fig. \ref{fig:RV}, we compare our RV values from the second half of 2021 with RV data published earlier by large spectroscopic surveys. J04034842+1551272 and J16564223-2108420 were both observed by APOGEE and GALAH (the corresponding IDs are reported in Table \ref{tab:AP}), and we included the RV values found in their catalogues in Fig. \ref{fig:RV}. We queried the data using the VizieR catalogue access tool \citep{10.26093/cds/vizier,vizier2000}. In the case of APOGEE, we used the seventeenth Data Release catalogue \citep{APOGEE-DR17}, which has the individual radial-velocity measurements as a main data product and includes a new radial velocity analysis performed with \texttt{Doppler} (\citealt{Nidever-doppler}; see \citealt{APOGEE-DR17} for further details). In the case of GALAH, we used the Third Data Release catalogue \citep{GALAH-DR3}, for which a new and improved pipeline was implemented \citep{Kos2016} with respect to previous releases. Figure \ref{fig:RV} shows all the radial velocity data as a function of time. The 1$\sigma$-uncertainties, as given in the catalogues or as we calculated, are included in the plot. In most cases they are smaller than the symbols. The data cover more than 2500 days in both cases. Even though it is far from enough to conclude anything about the orbital periods of these systems, it is enough to claim that both stars are RV-variable and, hence, very likely binary systems.

The variability is significant considering the data uncertainties in both cases, but the amplitude of the variability is smaller in the case of J16564223-2108420. Combining data from different sources is never trivial, because small offsets between the various instruments could lead to confusion. However, \cite{APOGEE-RVstds} cross-matched the APOGEE DR17 RV standard stars with the GALAH DR3 catalogue, and found 1839 stars in common. The authors measured the difference between the APOGEE and the GALAH RV ($\Delta RV$ = $RV_{\rm APOGEE} - RV_{\rm GALAH}$) for these 1839 stars and obtained a mean value for $\Delta RV$ of -0.031~$\mathrm{km s}^{-1}$ and a standard deviation of 0.299~$\mathrm{km s}^{-1}$. The $\Delta RV$ for J16564223-2108420 is around 1.5~$\mathrm{km s}^{-1}$, while less than 1\% of the standard stars from \cite{APOGEE-RVstds} present such high $\Delta RV$ values. This likely indicates that the observed variability in the case of J16564223-2108420 is also intrinsic and not instrumental. To shed light on this issue, we will apply for additional data to constrain both orbits in the future.

\begin{figure}[t]
\centering
 \includegraphics[width=0.5\textwidth]{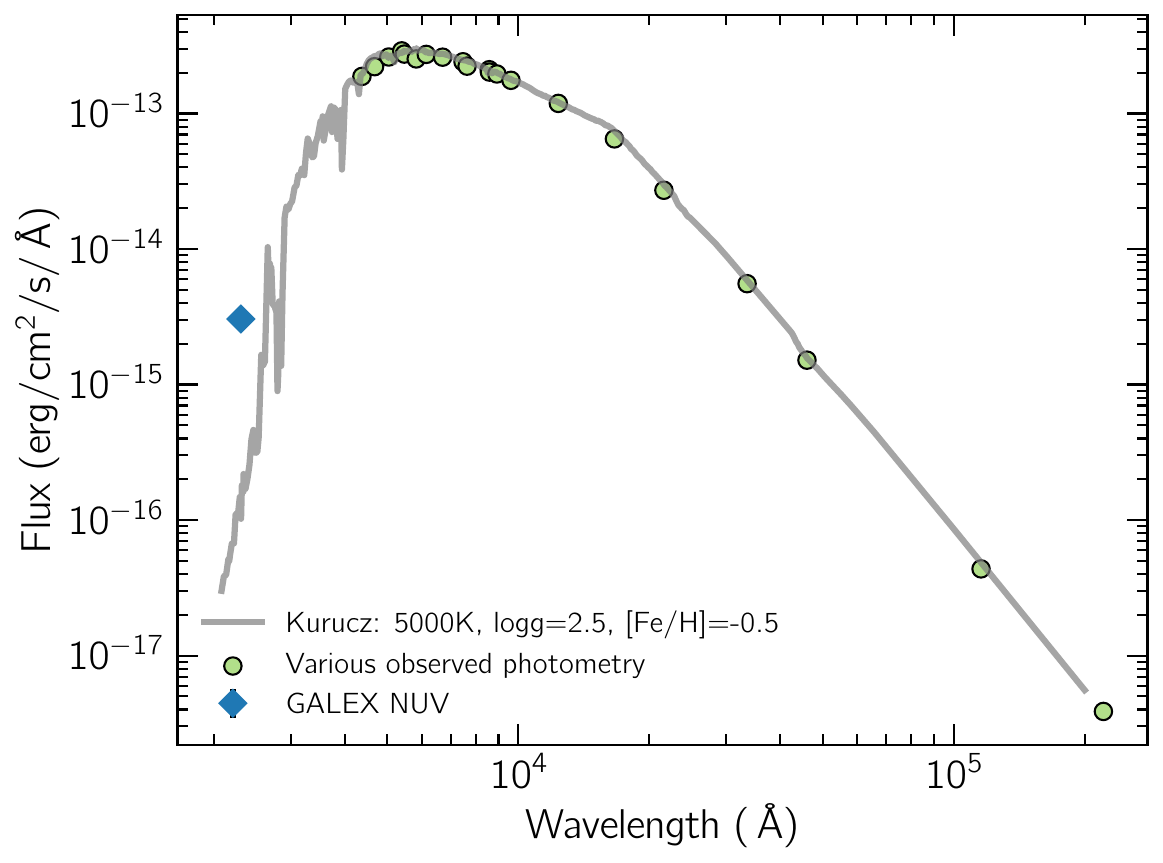} 
 \caption{Spectral energy distribution for J04034842+1551272 built with VOSA \citep{VOSA}. The photometry includes GALEX NUV \citep{GALEX2017}, Johnson B, V, R and I \citep{Hog2000, Henden2015}, SLOAN/SDSS g, r, i and z \citep{SDSS}, Gaia G, G$_{BP}$, and G$_{RP}$ \citep{GaiaSynthPhot}, 2MASS J, H, and K$_{S}$ \citep{2MASS}, and WISE W1, W2, W3 and W4. The model comes from the New Grids of ATLAS9 Model Atmospheres from \cite{Kurucz}.}
 \label{fig:SED}
\end{figure}

\subsubsection{The UV excess in 2MASS\,J04034842+1551272}\label{sect:SED}

Ultraviolet photometric excess has been observed in the Spectral Energy Distribution (SED) of Ba stars at different evolutionary stages \citep[e.g.][]{Bohm-Vitense00,Gray11, Shetye20} and is associated with a WD companion. We used the VOSA tool (Virtual Observatory SED Analyser; \citealt{VOSA}) to query broadband photometry available for our targets and build an SED. J16564223-2108420 did not have any GALEX far-UV (FUV) or near-UV (NUV) photometry. These filters, centred at 1528~\AA\ and 2271~\AA, respectively, are essential to detect UV excess and thereby argue the presence of a relatively hot ($>$10\,000~K) WD companion, so here we focus on J04034842+1551272.

Figure \ref{fig:SED} shows the available photometry used for this exercise, including the GALEX NUV data point from \cite{GALEX2017}, which appears about an order of magnitude brighter than the red giant best-fitting model. The photometry includes the following measurements: Johnson B, V, R and I from \cite{Hog2000, Henden2015}; SLOAN/SDSS g, r, i and z from \cite{SDSS}; Gaia G, G$_{BP}$, and G$_{RP}$ from \cite{GaiaSynthPhot}; 2MASS J, H, and K$_{S}$ from \cite{2MASS}; and WISE W1, W2, W3 and W4 from \cite{WISE}.	In order to evaluate the UV excess, we searched for a good atmospheric model that could fit the photometric data in the New Grids of ATLAS9 Model Atmospheres \citep{Kurucz}. We fixed $\log\,g$ to the closest value available in the grid to our seismic $\log\,g$, which is 2.5 dex. Then we allowed the effective temperature and the metallicity to vary between 4750 and 5250~K and between 0.0 and -1.0~dex, respectively.

The remaining free parameter is the extinction, $A_{\rm V}$. We found a value of $A_{\rm V}$ = 1.061 in \cite{GaiaEDR3summary} and $A_{\rm V}$ = 1.115 in the maps by \cite{Gontcharov2012}. Since the stellar parameters for J04034842+1551272 are well determined, we allowed $A_{\rm V}$ to vary in VOSA between 0.5 and 1.5 to avoid overfitting. Even though they are in the figure and the model reproduces them well, we did not include the WISE data points in the fit, in case there was a small infrared excess. The final best-fitting values are  $\teff=$~5000~K, $\log\,g=$~2.5~dex, [Fe/H]~=~-0.5 and $A_{\rm V}$ = 1.026 in great agreement with our spectroscopically determined values. VOSA does not interpolate between grid points, and this is the closest model in the grid by \cite{Kurucz} to our values (Table \ref{tab:AP}). Figure \ref{fig:SED} includes the model and the observed photometry, dereddened using the extinction law by \cite{Fitzpatrick99} and improved by \cite{Indebetouw05} in the infrared. It is clear from the figure that while most of the photometric data is very well reproduced by the best-fitting model taking the atmospheric extinction into account, the GALEX NUV point is brighter than expected for a red giant with the parameters derived for J04034842+1551272. This provides additional evidence suggesting that this target is a binary and that the companion is a WD, as expected for Ba stars.

\section{Summary and conclusions}\label{sect:conclusions}

This paper presents a detailed chemical analysis of two bright red giants that we classify as Ba stars for the first time. We used high-resolution UVES spectra and the public spectral software \texttt{iSpec}. J04034842+1551272 is a He-burning red clump star, while J16564223-2108420 is on the red giant branch. The two targets are strongly enhanced in Ba and other s-process elements, while they only have a mild enhancement in the r-process element europium. This allows us to conclude that the heavy metals in these stars have a dominant s-process origin, without a major contribution from the r-process. The heavy element abundances derived in this work are in agreement with the majority of Ba stars found in the literature.

Here we also show, for the first time, direct evidence of the binary nature of both targets. In the case of J04034842+1551272, the ultraviolet excess present in its spectral energy distribution suggests that its faint companion is a white dwarf. This is expected for Ba and related stars. The available radial velocity data for this target also shows significant variability, although the data is not enough to estimate its orbital period.

There are no UV measurements for J16564223-2108420 and its radial velocity data is not as strongly variable as in the case of the red clump star. However, it is still significant. Ba stars are known to have longer orbital periods, and long-term monitoring programs are often needed to determine their orbital properties. Hopefully, future releases of astrometric data combined with additional RV points will allow us to estimate the orbital parameters of these systems and the companion masses and use these targets to learn about mass-transfer and AGB nucleosynthesis models.

With this paper, two more well-studied Ba giants are added to the already-known sample of Ba stars. We presented a precise full abundance study based on high-resolution high-SNR data. Additionally, we flagged two bright binary stars previously undetected. This is a small step forward in the field that needs additional observational constraints to increase our knowledge of nucleosynthesis and binary interaction models.\\

\section*{acknowledgements}
\footnotesize{\textit{Acknowledgements}: This project received the support of a fellowship from "La Caixa" Foundation (ID 100010434). The fellowship code is LCF/BQ/PI23/11970031. SV thanks ANID (Beca Doctorado Nacional, folio 21220489) and Universidad Diego Portales for the financial support provided. SV and PJ acknowledge the Millennium Nucleus ERIS (ERIS NCN2021017) and FONDECYT (Regular number 1231057) for the funding. We want to thank the anonymous referee for their helpful and insightful comments.

This research has made use of the VizieR catalogue access tool, CDS, Strasbourg, France \citep{10.26093/cds/vizier}. The original description of the VizieR service was published in \citet{vizier2000}. This publication makes use of VOSA, developed under the Spanish Virtual Observatory (\url{https://svo.cab.inta-csic.es}) project funded by MCIN/AEI/10.13039/501100011033/ through grant PID2020-112949GB-I00. This work has made use of data from the European Space Agency (ESA) mission
{\it Gaia} (\url{https://www.cosmos.esa.int/gaia}), processed by the {\it Gaia} Data Processing and Analysis Consortium (DPAC, \url{https://www.cosmos.esa.int/web/gaia/dpac/consortium}). Funding for the DPAC has been provided by national institutions, in particular the institutions
participating in the {\it Gaia} Multilateral Agreement.}

\bibliography{reference-list.bib}{}

\begin{thebibliography}{}
\expandafter\ifx\csname natexlab\endcsname\relax\def\natexlab#1{#1}\fi
\providecommand{\url}[1]{\href{#1}{#1}}
\providecommand{\dodoi}[1]{doi:~\href{http://doi.org/#1}{\nolinkurl{#1}}}
\providecommand{\doeprint}[1]{\href{http://ascl.net/#1}{\nolinkurl{http://ascl.net/#1}}}
\providecommand{\doarXiv}[1]{\href{https://arxiv.org/abs/#1}{\nolinkurl{https://arxiv.org/abs/#1}}}

\bibitem[{{Abate} {et~al.}(2018){Abate}, {Pols}, \& {Stancliffe}}]{Abate18}
{Abate}, C., {Pols}, O.~R., \& {Stancliffe}, R.~J. 2018, \aap, 620, A63, \dodoi{10.1051/0004-6361/201833780}

\bibitem[{{Abdelkawy} {et~al.}(2017){Abdelkawy}, {Shaltout}, {Beheary}, \& {Bakry}}]{2017Abdelkawy}
{Abdelkawy}, A. G.~A., {Shaltout}, A. M.~K., {Beheary}, M.~M., \& {Bakry}, A. 2017, \mnras, 470, 4007, \dodoi{10.1093/mnras/stx1407}

\bibitem[{{Abdurro'uf} {et~al.}(2022){Abdurro'uf}, {Accetta}, {Aerts}, {Silva Aguirre}, {Ahumada}, {Ajgaonkar}, {Filiz Ak}, {Alam}, {Allende Prieto}, {Almeida}, {Anders}, {Anderson}, {Andrews}, {Anguiano}, {Aquino-Ort{\'\i}z}, {Arag{\'o}n-Salamanca}, {Argudo-Fern{\'a}ndez}, {Ata}, {Aubert}, {Avila-Reese}, {Badenes}, {Barb{\'a}}, {Barger}, {Barrera-Ballesteros}, {Beaton}, {Beers}, {Belfiore}, {Bender}, {Bernardi}, {Bershady}, {Beutler}, {Bidin}, {Bird}, {Bizyaev}, {Blanc}, {Blanton}, {Boardman}, {Bolton}, {Boquien}, {Borissova}, {Bovy}, {Brandt}, {Brown}, {Brownstein}, {Brusa}, {Buchner}, {Bundy}, {Burchett}, {Bureau}, {Burgasser}, {Cabang}, {Campbell}, {Cappellari}, {Carlberg}, {Wanderley}, {Carrera}, {Cash}, {Chen}, {Chen}, {Cherinka}, {Chiappini}, {Choi}, {Chojnowski}, {Chung}, {Clerc}, {Cohen}, {Comerford}, {Comparat}, {da Costa}, {Covey}, {Crane}, {Cruz-Gonzalez}, {Culhane}, {Cunha}, {Dai}, {Damke}, {Darling}, {Davidson}, {Davies}, {Dawson}, {De Lee}, {Diamond-Stanic}, {Cano-D{\'\i}az}, {S{\'a}nchez},
  {Donor}, {Duckworth}, {Dwelly}, {Eisenstein}, {Elsworth}, {Emsellem}, {Eracleous}, {Escoffier}, {Fan}, {Farr}, {Feng}, {Fern{\'a}ndez-Trincado}, {Feuillet}, {Filipp}, {Fillingham}, {Frinchaboy}, {Fromenteau}, {Galbany}, {Garc{\'\i}a}, {Garc{\'\i}a-Hern{\'a}ndez}, {Ge}, {Geisler}, {Gelfand}, {G{\'e}ron}, {Gibson}, {Goddy}, {Godoy-Rivera}, {Grabowski}, {Green}, {Greener}, {Grier}, {Griffith}, {Guo}, {Guy}, {Hadjara}, {Harding}, {Hasselquist}, {Hayes}, {Hearty}, {Hern{\'a}ndez}, {Hill}, {Hogg}, {Holtzman}, {Horta}, {Hsieh}, {Hsu}, {Hsu}, {Huber}, {Huertas-Company}, {Hutchinson}, {Hwang}, {Ibarra-Medel}, {Chitham}, {Ilha}, {Imig}, {Jaekle}, {Jayasinghe}, {Ji}, {Johnson}, {Jones}, {J{\"o}nsson}, {Katkov}, {Khalatyan}, {Kinemuchi}, {Kisku}, {Knapen}, {Kneib}, {Kollmeier}, {Kong}, {Kounkel}, {Kreckel}, {Krishnarao}, {Lacerna}, {Lane}, {Langgin}, {Lavender}, {Law}, {Lazarz}, {Leung}, {Leung}, {Lewis}, {Li}, {Li}, {Lian}, {Liang}, {Lin}, {Lin}, {Lin}, {Lintott}, {Long}, {Longa-Pe{\~n}a}, {L{\'o}pez-Cob{\'a}}, {Lu},
  {Lundgren}, {Luo}, {Mackereth}, {de la Macorra}, {Mahadevan}, {Majewski}, {Manchado}, {Mandeville}, {Maraston}, {Margalef-Bentabol}, {Masseron}, {Masters}, {Mathur}, {McDermid}, {Mckay}, {Merloni}, {Merrifield}, {Meszaros}, {Miglio}, {Di Mille}, {Minniti}, {Minsley}, {Monachesi}, {Moon}, {Mosser}, {Mulchaey}, {Muna}, {Mu{\~n}oz}, {Myers}, {Myers}, {Nadathur}, {Nair}, {Nandra}, {Neumann}, {Newman}, {Nidever}, {Nikakhtar}, {Nitschelm}, {O'Connell}, {Garma-Oehmichen}, {Luan Souza de Oliveira}, {Olney}, {Oravetz}, {Ortigoza-Urdaneta}, {Osorio}, {Otter}, {Pace}, {Padilla}, {Pan}, {Pan}, {Parikh}, {Parker}, {Peirani}, {Pe{\~n}a Ram{\'\i}rez}, {Penny}, {Percival}, {Perez-Fournon}, {Pinsonneault}, {Poidevin}, {Poovelil}, {Price-Whelan}, {B{\'a}rbara de Andrade Queiroz}, {Raddick}, {Ray}, {Rembold}, {Riddle}, {Riffel}, {Riffel}, {Rix}, {Robin}, {Rodr{\'\i}guez-Puebla}, {Roman-Lopes}, {Rom{\'a}n-Z{\'u}{\~n}iga}, {Rose}, {Ross}, {Rossi}, {Rubin}, {Salvato}, {S{\'a}nchez}, {S{\'a}nchez-Gallego}, {Sanderson}, {Santana
  Rojas}, {Sarceno}, {Sarmiento}, {Sayres}, {Sazonova}, {Schaefer}, {Schiavon}, {Schlegel}, {Schneider}, {Schultheis}, {Schwope}, {Serenelli}, {Serna}, {Shao}, {Shapiro}, {Sharma}, {Shen}, {Shetrone}, {Shu}, {Simon}, {Skrutskie}, {Smethurst}, {Smith}, {Sobeck}, {Spoo}, {Sprague}, {Stark}, {Stassun}, {Steinmetz}, {Stello}, {Stone-Martinez}, {Storchi-Bergmann}, {Stringfellow}, {Stutz}, {Su}, {Taghizadeh-Popp}, {Talbot}, {Tayar}, {Telles}, {Teske}, {Thakar}, {Theissen}, {Tkachenko}, {Thomas}, {Tojeiro}, {Hernandez Toledo}, {Troup}, {Trump}, {Trussler}, {Turner}, {Tuttle}, {Unda-Sanzana}, {V{\'a}zquez-Mata}, {Valentini}, {Valenzuela}, {Vargas-Gonz{\'a}lez}, {Vargas-Maga{\~n}a}, {Alfaro}, {Villanova}, {Vincenzo}, {Wake}, {Warfield}, {Washington}, {Weaver}, {Weijmans}, {Weinberg}, {Weiss}, {Westfall}, {Wild}, {Wilde}, {Wilson}, {Wilson}, {Wilson}, {Wolf}, {Wood-Vasey}, {Yan}, {Zamora}, {Zasowski}, {Zhang}, {Zhao}, {Zheng}, {Zheng}, \& {Zhu}}]{APOGEE-DR17}
{Abdurro'uf}, {Accetta}, K., {Aerts}, C., {et~al.} 2022, \apjs, 259, 35, \dodoi{10.3847/1538-4365/ac4414}

\bibitem[{{Alam} {et~al.}(2015){Alam}, {Albareti}, {Allende Prieto}, {Anders}, {Anderson}, {Anderton}, {Andrews}, {Armengaud}, {Aubourg}, {Bailey}, {Basu}, {Bautista}, {Beaton}, {Beers}, {Bender}, {Berlind}, {Beutler}, {Bhardwaj}, {Bird}, {Bizyaev}, {Blake}, {Blanton}, {Blomqvist}, {Bochanski}, {Bolton}, {Bovy}, {Shelden Bradley}, {Brandt}, {Brauer}, {Brinkmann}, {Brown}, {Brownstein}, {Burden}, {Burtin}, {Busca}, {Cai}, {Capozzi}, {Carnero Rosell}, {Carr}, {Carrera}, {Chambers}, {Chaplin}, {Chen}, {Chiappini}, {Chojnowski}, {Chuang}, {Clerc}, {Comparat}, {Covey}, {Croft}, {Cuesta}, {Cunha}, {da Costa}, {Da Rio}, {Davenport}, {Dawson}, {De Lee}, {Delubac}, {Deshpande}, {Dhital}, {Dutra-Ferreira}, {Dwelly}, {Ealet}, {Ebelke}, {Edmondson}, {Eisenstein}, {Ellsworth}, {Elsworth}, {Epstein}, {Eracleous}, {Escoffier}, {Esposito}, {Evans}, {Fan}, {Fern{\'a}ndez-Alvar}, {Feuillet}, {Filiz Ak}, {Finley}, {Finoguenov}, {Flaherty}, {Fleming}, {Font-Ribera}, {Foster}, {Frinchaboy}, {Galbraith-Frew}, {Garc{\'\i}a},
  {Garc{\'\i}a-Hern{\'a}ndez}, {Garc{\'\i}a P{\'e}rez}, {Gaulme}, {Ge}, {G{\'e}nova-Santos}, {Georgakakis}, {Ghezzi}, {Gillespie}, {Girardi}, {Goddard}, {Gontcho}, {Gonz{\'a}lez Hern{\'a}ndez}, {Grebel}, {Green}, {Grieb}, {Grieves}, {Gunn}, {Guo}, {Harding}, {Hasselquist}, {Hawley}, {Hayden}, {Hearty}, {Hekker}, {Ho}, {Hogg}, {Holley-Bockelmann}, {Holtzman}, {Honscheid}, {Huber}, {Huehnerhoff}, {Ivans}, {Jiang}, {Johnson}, {Kinemuchi}, {Kirkby}, {Kitaura}, {Klaene}, {Knapp}, {Kneib}, {Koenig}, {Lam}, {Lan}, {Lang}, {Laurent}, {Le Goff}, {Leauthaud}, {Lee}, {Lee}, {Licquia}, {Liu}, {Long}, {L{\'o}pez-Corredoira}, {Lorenzo-Oliveira}, {Lucatello}, {Lundgren}, {Lupton}, {Mack}, {Mahadevan}, {Maia}, {Majewski}, {Malanushenko}, {Malanushenko}, {Manchado}, {Manera}, {Mao}, {Maraston}, {Marchwinski}, {Margala}, {Martell}, {Martig}, {Masters}, {Mathur}, {McBride}, {McGehee}, {McGreer}, {McMahon}, {M{\'e}nard}, {Menzel}, {Merloni}, {M{\'e}sz{\'a}ros}, {Miller}, {Miralda-Escud{\'e}}, {Miyatake}, {Montero-Dorta}, {More},
  {Morganson}, {Morice-Atkinson}, {Morrison}, {Mosser}, {Muna}, {Myers}, {Nandra}, {Newman}, {Neyrinck}, {Nguyen}, {Nichol}, {Nidever}, {Noterdaeme}, {Nuza}, {O'Connell}, {O'Connell}, {O'Connell}, {Ogando}, {Olmstead}, {Oravetz}, {Oravetz}, {Osumi}, {Owen}, {Padgett}, {Padmanabhan}, {Paegert}, {Palanque-Delabrouille}, {Pan}, {Parejko}, {P{\^a}ris}, {Park}, {Pattarakijwanich}, {Pellejero-Ibanez}, {Pepper}, {Percival}, {P{\'e}rez-Fournon}, {P{\'e}rez-R{\`a}fols}, {Petitjean}, {Pieri}, {Pinsonneault}, {Porto de Mello}, {Prada}, {Prakash}, {Price-Whelan}, {Protopapas}, {Raddick}, {Rahman}, {Reid}, {Rich}, {Rix}, {Robin}, {Rockosi}, {Rodrigues}, {Rodr{\'\i}guez-Torres}, {Roe}, {Ross}, {Ross}, {Rossi}, {Ruan}, {Rubi{\~n}o-Mart{\'\i}n}, {Rykoff}, {Salazar-Albornoz}, {Salvato}, {Samushia}, {S{\'a}nchez}, {Santiago}, {Sayres}, {Schiavon}, {Schlegel}, {Schmidt}, {Schneider}, {Schultheis}, {Schwope}, {Sc{\'o}ccola}, {Scott}, {Sellgren}, {Seo}, {Serenelli}, {Shane}, {Shen}, {Shetrone}, {Shu}, {Silva Aguirre}, {Sivarani},
  {Skrutskie}, {Slosar}, {Smith}, {Sobreira}, {Souto}, {Stassun}, {Steinmetz}, {Stello}, {Strauss}, {Streblyanska}, {Suzuki}, {Swanson}, {Tan}, {Tayar}, {Terrien}, {Thakar}, {Thomas}, {Thomas}, {Thompson}, {Tinker}, {Tojeiro}, {Troup}, {Vargas-Maga{\~n}a}, {Vazquez}, {Verde}, {Viel}, {Vogt}, {Wake}, {Wang}, {Weaver}, {Weinberg}, {Weiner}, {White}, {Wilson}, {Wisniewski}, {Wood-Vasey}, {Ye`che}, {York}, {Zakamska}, {Zamora}, {Zasowski}, {Zehavi}, {Zhao}, {Zheng}, {Zhou}, {Zhou}, {Zou}, \& {Zhu}}]{SDSS}
{Alam}, S., {Albareti}, F.~D., {Allende Prieto}, C., {et~al.} 2015, \apjs, 219, 12, \dodoi{10.1088/0067-0049/219/1/12}

\bibitem[{{Alexeeva} {et~al.}(2023){Alexeeva}, {Wang}, {Zhao}, {Wang}, {Wu}, {Wang}, {Yan}, \& {Shi}}]{2023Alexeeva}
{Alexeeva}, S., {Wang}, Y., {Zhao}, G., {et~al.} 2023, \apj, 957, 10, \dodoi{10.3847/1538-4357/acf5e1}

\bibitem[{{Allen} \& {Barbuy}(2006)}]{2006Allen}
{Allen}, D.~M., \& {Barbuy}, B. 2006, \aap, 454, 917, \dodoi{10.1051/0004-6361:20064968}

\bibitem[{{Alvarez} \& {Plez}(1998)}]{1998A&A...330.1109A}
{Alvarez}, R., \& {Plez}, B. 1998, \aap, 330, 1109, \dodoi{10.48550/arXiv.astro-ph/9710157}

\bibitem[{{Andrievsky} {et~al.}(2009){Andrievsky}, {Spite}, {Korotin}, {Spite}, {Fran{\c{c}}ois}, {Bonifacio}, {Cayrel}, \& {Hill}}]{2009Andrievsky}
{Andrievsky}, S.~M., {Spite}, M., {Korotin}, S.~A., {et~al.} 2009, \aap, 494, 1083, \dodoi{10.1051/0004-6361:200810894}

\bibitem[{{Ballester} {et~al.}(2000){Ballester}, {Modigliani}, {Boitquin}, {Cristiani}, {Hanuschik}, {Kaufer}, \& {Wolf}}]{UVESpipeline}
{Ballester}, P., {Modigliani}, A., {Boitquin}, O., {et~al.} 2000, The Messenger, 101, 31

\bibitem[{{Bayo} {et~al.}(2008){Bayo}, {Rodrigo}, {Barrado Y Navascu{\'e}s}, {Solano}, {Guti{\'e}rrez}, {Morales-Calder{\'o}n}, \& {Allard}}]{VOSA}
{Bayo}, A., {Rodrigo}, C., {Barrado Y Navascu{\'e}s}, D., {et~al.} 2008, \aap, 492, 277, \dodoi{10.1051/0004-6361:200810395}

\bibitem[{{Beers} \& {Christlieb}(2005)}]{BeersChristlieb05}
{Beers}, T.~C., \& {Christlieb}, N. 2005, \araa, 43, 531, \dodoi{10.1146/annurev.astro.42.053102.134057}

\bibitem[{{Bergemann} {et~al.}(2012){Bergemann}, {Hansen}, {Bautista}, \& {Ruchti}}]{2012Bergemann}
{Bergemann}, M., {Hansen}, C.~J., {Bautista}, M., \& {Ruchti}, G. 2012, \aap, 546, A90, \dodoi{10.1051/0004-6361/201219406}

\bibitem[{{Bianchi} {et~al.}(2017){Bianchi}, {Shiao}, \& {Thilker}}]{GALEX2017}
{Bianchi}, L., {Shiao}, B., \& {Thilker}, D. 2017, \apjs, 230, 24, \dodoi{10.3847/1538-4365/aa7053}

\bibitem[{{Bidelman} \& {Keenan}(1951)}]{BidelmanKeenan51}
{Bidelman}, W.~P., \& {Keenan}, P.~C. 1951, ApJ, 114, 473, \dodoi{10.1086/145488}

\bibitem[{{Bisterzo} {et~al.}(2006){Bisterzo}, {Gallino}, {Straniero}, {Ivans}, {K{\"a}ppeler}, \& {Aoki}}]{2006MmBisterzo}
{Bisterzo}, S., {Gallino}, R., {Straniero}, O., {et~al.} 2006, \memsai, 77, 985

\bibitem[{{Bisterzo} {et~al.}(2014){Bisterzo}, {Travaglio}, {Gallino}, {Wiescher}, \& {K{\"a}ppeler}}]{2014Bisterzo}
{Bisterzo}, S., {Travaglio}, C., {Gallino}, R., {Wiescher}, M., \& {K{\"a}ppeler}, F. 2014, \apj, 787, 10, \dodoi{10.1088/0004-637X/787/1/10}

\bibitem[{{Bisterzo} {et~al.}(2015){Bisterzo}, {Gallino}, {K{\"a}ppeler}, {Wiescher}, {Imbriani}, {Straniero}, {Cristallo}, {G{\"o}rres}, \& {deBoer}}]{2015Bisterzo}
{Bisterzo}, S., {Gallino}, R., {K{\"a}ppeler}, F., {et~al.} 2015, \mnras, 449, 506, \dodoi{10.1093/mnras/stv271}

\bibitem[{{Blanco-Cuaresma}(2019)}]{2019Blanco}
{Blanco-Cuaresma}, S. 2019, \mnras, 486, 2075, \dodoi{10.1093/mnras/stz549}

\bibitem[{{Blanco-Cuaresma} {et~al.}(2014{\natexlab{a}}){Blanco-Cuaresma}, {Soubiran}, {Heiter}, \& {Jofr{\'e}}}]{2014Blanco}
{Blanco-Cuaresma}, S., {Soubiran}, C., {Heiter}, U., \& {Jofr{\'e}}, P. 2014{\natexlab{a}}, \aap, 569, A111, \dodoi{10.1051/0004-6361/201423945}

\bibitem[{{Blanco-Cuaresma} {et~al.}(2014{\natexlab{b}}){Blanco-Cuaresma}, {Soubiran}, {Jofr{\'e}}, \& {Heiter}}]{Blanco-Cuaresma2014}
{Blanco-Cuaresma}, S., {Soubiran}, C., {Jofr{\'e}}, P., \& {Heiter}, U. 2014{\natexlab{b}}, \aap, 566, A98, \dodoi{10.1051/0004-6361/201323153}

\bibitem[{{B{\"o}hm-Vitense} {et~al.}(2000){B{\"o}hm-Vitense}, {Carpenter}, {Robinson}, {Ake}, \& {Brown}}]{Bohm-Vitense00}
{B{\"o}hm-Vitense}, E., {Carpenter}, K., {Robinson}, R., {Ake}, T., \& {Brown}, J. 2000, \apj, 533, 969, \dodoi{10.1086/308678}

\bibitem[{{B{\"o}hm-Vitense} {et~al.}(1984){B{\"o}hm-Vitense}, {Nemec}, \& {Proffitt}}]{Bohm-Vitense84}
{B{\"o}hm-Vitense}, E., {Nemec}, J., \& {Proffitt}, C. 1984, \apj, 278, 726, \dodoi{10.1086/161843}

\bibitem[{{Brogaard} {et~al.}(2016){Brogaard}, {Jessen-Hansen}, {Handberg}, {Arentoft}, {Frandsen}, {Grundahl}, {Bruntt}, {Sandquist}, {Miglio}, {Beck}, {Thygesen}, {Kj{\ae}rgaard}, \& {Haugaard}}]{2016Brogaard}
{Brogaard}, K., {Jessen-Hansen}, J., {Handberg}, R., {et~al.} 2016, Astronomische Nachrichten, 337, 793, \dodoi{10.1002/asna.201612374}

\bibitem[{{Brown} {et~al.}(1991){Brown}, {Gilliland}, {Noyes}, \& {Ramsey}}]{1991Brown}
{Brown}, T.~M., {Gilliland}, R.~L., {Noyes}, R.~W., \& {Ramsey}, L.~W. 1991, \apj, 368, 599, \dodoi{10.1086/169725}

\bibitem[{{Buder} {et~al.}(2021){Buder}, {Sharma}, {Kos}, {Amarsi}, {Nordlander}, {Lind}, {Martell}, {Asplund}, {Bland-Hawthorn}, {Casey}, {de Silva}, {D'Orazi}, {Freeman}, {Hayden}, {Lewis}, {Lin}, {Schlesinger}, {Simpson}, {Stello}, {Zucker}, {Zwitter}, {Beeson}, {Buck}, {Casagrande}, {Clark}, {{\v{C}}otar}, {da Costa}, {de Grijs}, {Feuillet}, {Horner}, {Kafle}, {Khanna}, {Kobayashi}, {Liu}, {Montet}, {Nandakumar}, {Nataf}, {Ness}, {Spina}, {Tepper-Garc{\'\i}a}, {Ting}, {Traven}, {Vogrin{\v{c}}i{\v{c}}}, {Wittenmyer}, {Wyse}, {{\v{Z}}erjal}, \& {Galah Collaboration}}]{GALAH-DR3}
{Buder}, S., {Sharma}, S., {Kos}, J., {et~al.} 2021, \mnras, 506, 150, \dodoi{10.1093/mnras/stab1242}

\bibitem[{{Burbidge} {et~al.}(1957){Burbidge}, {Burbidge}, {Fowler}, \& {Hoyle}}]{Burbidge57}
{Burbidge}, E.~M., {Burbidge}, G.~R., {Fowler}, W.~A., \& {Hoyle}, F. 1957, Reviews of Modern Physics, 29, 547, \dodoi{10.1103/RevModPhys.29.547}

\bibitem[{{Casamiquela} {et~al.}(2020){Casamiquela}, {Tarricq}, {Soubiran}, {Blanco-Cuaresma}, {Jofr{\'e}}, {Heiter}, \& {Tucci Maia}}]{2020Casamiquela}
{Casamiquela}, L., {Tarricq}, Y., {Soubiran}, C., {et~al.} 2020, \aap, 635, A8, \dodoi{10.1051/0004-6361/201936978}

\bibitem[{{Castelli} \& {Kurucz}(2003)}]{Kurucz}
{Castelli}, F., \& {Kurucz}, R.~L. 2003, in Modelling of Stellar Atmospheres, ed. N.~{Piskunov}, W.~W. {Weiss}, \& D.~F. {Gray}, Vol. 210, A20.
\newblock \doarXiv{astro-ph/0405087}

\bibitem[{{Cescutti} {et~al.}(2021){Cescutti}, {Morossi}, {Franchini}, {Di Marcantonio}, {Chiappini}, {Steffen}, {Valentini}, {Fran{\c{c}}ois}, {Christlieb}, {Cort{\'e}s}, {Kobayashi}, \& {Depagne}}]{2021Cescutti}
{Cescutti}, G., {Morossi}, C., {Franchini}, M., {et~al.} 2021, \aap, 654, A164, \dodoi{10.1051/0004-6361/202141355}

\bibitem[{{Choplin} {et~al.}(2024){Choplin}, {Siess}, {Goriely}, \& {Martinet}}]{Choplin24}
{Choplin}, A., {Siess}, L., {Goriely}, S., \& {Martinet}, S. 2024, \aap, 684, A206, \dodoi{10.1051/0004-6361/202348957}

\bibitem[{{Cowan} \& {Rose}(1977)}]{i-process}
{Cowan}, J.~J., \& {Rose}, W.~K. 1977, \apj, 212, 149, \dodoi{10.1086/155030}

\bibitem[{{Creevey} {et~al.}(2013){Creevey}, {Th{\'e}venin}, {Basu}, {Chaplin}, {Bigot}, {Elsworth}, {Huber}, {Monteiro}, \& {Serenelli}}]{2013Creevey}
{Creevey}, O.~L., {Th{\'e}venin}, F., {Basu}, S., {et~al.} 2013, \mnras, 431, 2419, \dodoi{10.1093/mnras/stt336}

\bibitem[{{Cristallo} {et~al.}(2009{\natexlab{a}}){Cristallo}, {Straniero}, {Gallino}, {Piersanti}, {Dom{\'\i}nguez}, \& {Lederer}}]{Cristallo09}
{Cristallo}, S., {Straniero}, O., {Gallino}, R., {et~al.} 2009{\natexlab{a}}, \apj, 696, 797, \dodoi{10.1088/0004-637X/696/1/797}

\bibitem[{{Cristallo} {et~al.}(2009{\natexlab{b}}){Cristallo}, {Straniero}, {Gallino}, {Piersanti}, {Dom{\'\i}nguez}, \& {Lederer}}]{2009Cristallo}
---. 2009{\natexlab{b}}, \apj, 696, 797, \dodoi{10.1088/0004-637X/696/1/797}

\bibitem[{{Cristallo} {et~al.}(2015){Cristallo}, {Straniero}, {Piersanti}, \& {Gobrecht}}]{Cristallo15}
{Cristallo}, S., {Straniero}, O., {Piersanti}, L., \& {Gobrecht}, D. 2015, \apjs, 219, 40, \dodoi{10.1088/0067-0049/219/2/40}

\bibitem[{{Cseh} {et~al.}(2018){Cseh}, {Lugaro}, {D'Orazi}, {de Castro}, {Pereira}, {Karakas}, {Moln{\'a}r}, {Plachy}, {Szab{\'o}}, {Pignatari}, \& {Cristallo}}]{Cseh18}
{Cseh}, B., {Lugaro}, M., {D'Orazi}, V., {et~al.} 2018, \aap, 620, A146, \dodoi{10.1051/0004-6361/201834079}

\bibitem[{{Cseh} {et~al.}(2022){Cseh}, {Vil{\'a}gos}, {Roriz}, {Pereira}, {D'Orazi}, {Karakas}, {So{\'o}s}, {Drake}, {Junqueira}, \& {Lugaro}}]{Cseh22}
{Cseh}, B., {Vil{\'a}gos}, B., {Roriz}, M.~P., {et~al.} 2022, \aap, 660, A128, \dodoi{10.1051/0004-6361/202142468}

\bibitem[{{Cui} {et~al.}(2014){Cui}, {Zhang}, {Shi}, {Zhao}, {Wang}, \& {Niu}}]{2014Cui}
{Cui}, W.~Y., {Zhang}, B., {Shi}, J.~R., {et~al.} 2014, \aap, 566, A16, \dodoi{10.1051/0004-6361/201323295}

\bibitem[{{de Castro} {et~al.}(2016){de Castro}, {Pereira}, {Roig}, {Jilinski}, {Drake}, {Chavero}, \& {Sales Silva}}]{deCastro16}
{de Castro}, D.~B., {Pereira}, C.~B., {Roig}, F., {et~al.} 2016, \mnras, 459, 4299, \dodoi{10.1093/mnras/stw815}

\bibitem[{{De Silva} {et~al.}(2015){De Silva}, {Freeman}, {Bland-Hawthorn}, {Martell}, {de Boer}, {Asplund}, {Keller}, {Sharma}, {Zucker}, {Zwitter}, {Anguiano}, {Bacigalupo}, {Bayliss}, {Beavis}, {Bergemann}, {Campbell}, {Cannon}, {Carollo}, {Casagrande}, {Casey}, {Da Costa}, {D'Orazi}, {Dotter}, {Duong}, {Heger}, {Ireland}, {Kafle}, {Kos}, {Lattanzio}, {Lewis}, {Lin}, {Lind}, {Munari}, {Nataf}, {O'Toole}, {Parker}, {Reid}, {Schlesinger}, {Sheinis}, {Simpson}, {Stello}, {Ting}, {Traven}, {Watson}, {Wittenmyer}, {Yong}, \& {{\v{Z}}erjal}}]{GALAHsurvey}
{De Silva}, G.~M., {Freeman}, K.~C., {Bland-Hawthorn}, J., {et~al.} 2015, \mnras, 449, 2604, \dodoi{10.1093/mnras/stv327}

\bibitem[{{Dekker} {et~al.}(2000){Dekker}, {D'Odorico}, {Kaufer}, {Delabre}, \& {Kotzlowski}}]{UVES}
{Dekker}, H., {D'Odorico}, S., {Kaufer}, A., {Delabre}, B., \& {Kotzlowski}, H. 2000, in Society of Photo-Optical Instrumentation Engineers (SPIE) Conference Series, Vol. 4008, Optical and IR Telescope Instrumentation and Detectors, ed. M.~{Iye} \& A.~F. {Moorwood}, 534--545, \dodoi{10.1117/12.395512}

\bibitem[{{den Hartogh} {et~al.}(2023){den Hartogh}, {Yag{\"u}e L{\'o}pez}, {Cseh}, {Pignatari}, {Vil{\'a}gos}, {Roriz}, {Pereira}, {Drake}, {Junqueira}, \& {Lugaro}}]{denHartogh23}
{den Hartogh}, J.~W., {Yag{\"u}e L{\'o}pez}, A., {Cseh}, B., {et~al.} 2023, \aap, 672, A143, \dodoi{10.1051/0004-6361/202244189}

\bibitem[{{Eitner} {et~al.}(2019){Eitner}, {Bergemann}, \& {Larsen}}]{2019Eitner}
{Eitner}, P., {Bergemann}, M., \& {Larsen}, S. 2019, \aap, 627, A40, \dodoi{10.1051/0004-6361/201935416}

\bibitem[{{Escorza} \& {De Rosa}(2023)}]{EscorzaDeRosa23}
{Escorza}, A., \& {De Rosa}, R.~J. 2023, \aap, 671, A97, \dodoi{10.1051/0004-6361/202244782}

\bibitem[{{Escorza} {et~al.}(2020){Escorza}, {Siess}, {Van Winckel}, \& {Jorissen}}]{Escorza20}
{Escorza}, A., {Siess}, L., {Van Winckel}, H., \& {Jorissen}, A. 2020, \aap, 639, A24, \dodoi{10.1051/0004-6361/202037487}

\bibitem[{{Escorza} {et~al.}(2017){Escorza}, {Boffin}, {Jorissen}, {Van Eck}, {Siess}, {Van Winckel}, {Karinkuzhi}, {Shetye}, \& {Pourbaix}}]{Escorza17}
{Escorza}, A., {Boffin}, H.~M.~J., {Jorissen}, A., {et~al.} 2017, A\&A, 608, A100, \dodoi{10.1051/0004-6361/201731832}

\bibitem[{{Escorza} {et~al.}(2019){Escorza}, {Karinkuzhi}, {Jorissen}, {Siess}, {Van Winckel}, {Pourbaix}, {Johnston}, {Miszalski}, {Oomen}, {Abdul-Masih}, {Boffin}, {North}, {Manick}, {Shetye}, \& {Miko{\l}ajewska}}]{Escorza19}
{Escorza}, A., {Karinkuzhi}, D., {Jorissen}, A., {et~al.} 2019, \aap, 626, A128, \dodoi{10.1051/0004-6361/201935390}

\bibitem[{{Fishlock} {et~al.}(2014){Fishlock}, {Karakas}, {Lugaro}, \& {Yong}}]{2014Fishlock}
{Fishlock}, C.~K., {Karakas}, A.~I., {Lugaro}, M., \& {Yong}, D. 2014, \apj, 797, 44, \dodoi{10.1088/0004-637X/797/1/44}

\bibitem[{{Fitzpatrick}(1999)}]{Fitzpatrick99}
{Fitzpatrick}, E.~L. 1999, \pasp, 111, 63, \dodoi{10.1086/316293}

\bibitem[{{Freudling} {et~al.}(2013){Freudling}, {Romaniello}, {Bramich}, {Ballester}, {Forchi}, {Garc{\'{\i}}a-Dabl{\'o}}, {Moehler}, \& {Neeser}}]{2013Freudling}
{Freudling}, W., {Romaniello}, M., {Bramich}, D.~M., {et~al.} 2013, \aap, 559, A96, \dodoi{10.1051/0004-6361/201322494}

\bibitem[{{Gaia Collaboration} {et~al.}(2021){Gaia Collaboration}, {Brown}, {Vallenari}, {Prusti}, {de Bruijne}, {Babusiaux}, {Biermann}, {Creevey}, {Evans}, {Eyer}, {Hutton}, {Jansen}, {Jordi}, {Klioner}, {Lammers}, {Lindegren}, {Luri}, {Mignard}, {Panem}, {Pourbaix}, {Randich}, {Sartoretti}, {Soubiran}, {Walton}, {Arenou}, {Bailer-Jones}, {Bastian}, {Cropper}, {Drimmel}, {Katz}, {Lattanzi}, {van Leeuwen}, {Bakker}, {Cacciari}, {Casta{\~n}eda}, {De Angeli}, {Ducourant}, {Fabricius}, {Fouesneau}, {Fr{\'e}mat}, {Guerra}, {Guerrier}, {Guiraud}, {Jean-Antoine Piccolo}, {Masana}, {Messineo}, {Mowlavi}, {Nicolas}, {Nienartowicz}, {Pailler}, {Panuzzo}, {Riclet}, {Roux}, {Seabroke}, {Sordo}, {Tanga}, {Th{\'e}venin}, {Gracia-Abril}, {Portell}, {Teyssier}, {Altmann}, {Andrae}, {Bellas-Velidis}, {Benson}, {Berthier}, {Blomme}, {Brugaletta}, {Burgess}, {Busso}, {Carry}, {Cellino}, {Cheek}, {Clementini}, {Damerdji}, {Davidson}, {Delchambre}, {Dell'Oro}, {Fern{\'a}ndez-Hern{\'a}ndez}, {Galluccio}, {Garc{\'\i}a-Lario},
  {Garcia-Reinaldos}, {Gonz{\'a}lez-N{\'u}{\~n}ez}, {Gosset}, {Haigron}, {Halbwachs}, {Hambly}, {Harrison}, {Hatzidimitriou}, {Heiter}, {Hern{\'a}ndez}, {Hestroffer}, {Hodgkin}, {Holl}, {Jan{\ss}en}, {Jevardat de Fombelle}, {Jordan}, {Krone-Martins}, {Lanzafame}, {L{\"o}ffler}, {Lorca}, {Manteiga}, {Marchal}, {Marrese}, {Moitinho}, {Mora}, {Muinonen}, {Osborne}, {Pancino}, {Pauwels}, {Petit}, {Recio-Blanco}, {Richards}, {Riello}, {Rimoldini}, {Robin}, {Roegiers}, {Rybizki}, {Sarro}, {Siopis}, {Smith}, {Sozzetti}, {Ulla}, {Utrilla}, {van Leeuwen}, {van Reeven}, {Abbas}, {Abreu Aramburu}, {Accart}, {Aerts}, {Aguado}, {Ajaj}, {Altavilla}, {{\'A}lvarez}, {{\'A}lvarez Cid-Fuentes}, {Alves}, {Anderson}, {Anglada Varela}, {Antoja}, {Audard}, {Baines}, {Baker}, {Balaguer-N{\'u}{\~n}ez}, {Balbinot}, {Balog}, {Barache}, {Barbato}, {Barros}, {Barstow}, {Bartolom{\'e}}, {Bassilana}, {Bauchet}, {Baudesson-Stella}, {Becciani}, {Bellazzini}, {Bernet}, {Bertone}, {Bianchi}, {Blanco-Cuaresma}, {Boch}, {Bombrun}, {Bossini},
  {Bouquillon}, {Bragaglia}, {Bramante}, {Breedt}, {Bressan}, {Brouillet}, {Bucciarelli}, {Burlacu}, {Busonero}, {Butkevich}, {Buzzi}, {Caffau}, {Cancelliere}, {C{\'a}novas}, {Cantat-Gaudin}, {Carballo}, {Carlucci}, {Carnerero}, {Carrasco}, {Casamiquela}, {Castellani}, {Castro-Ginard}, {Castro Sampol}, {Chaoul}, {Charlot}, {Chemin}, {Chiavassa}, {Cioni}, {Comoretto}, {Cooper}, {Cornez}, {Cowell}, {Crifo}, {Crosta}, {Crowley}, {Dafonte}, {Dapergolas}, {David}, {David}, {de Laverny}, {De Luise}, {De March}, {De Ridder}, {de Souza}, {de Teodoro}, {de Torres}, {del Peloso}, {del Pozo}, {Delbo}, {Delgado}, {Delgado}, {Delisle}, {Di Matteo}, {Diakite}, {Diener}, {Distefano}, {Dolding}, {Eappachen}, {Edvardsson}, {Enke}, {Esquej}, {Fabre}, {Fabrizio}, {Faigler}, {Fedorets}, {Fernique}, {Fienga}, {Figueras}, {Fouron}, {Fragkoudi}, {Fraile}, {Franke}, {Gai}, {Garabato}, {Garcia-Gutierrez}, {Garc{\'\i}a-Torres}, {Garofalo}, {Gavras}, {Gerlach}, {Geyer}, {Giacobbe}, {Gilmore}, {Girona}, {Giuffrida}, {Gomel}, {Gomez},
  {Gonzalez-Santamaria}, {Gonz{\'a}lez-Vidal}, {Granvik}, {Guti{\'e}rrez-S{\'a}nchez}, {Guy}, {Hauser}, {Haywood}, {Helmi}, {Hidalgo}, {Hilger}, {H{\l}adczuk}, {Hobbs}, {Holland}, {Huckle}, {Jasniewicz}, {Jonker}, {Juaristi Campillo}, {Julbe}, {Karbevska}, {Kervella}, {Khanna}, {Kochoska}, {Kontizas}, {Kordopatis}, {Korn}, {Kostrzewa-Rutkowska}, {Kruszy{\'n}ska}, {Lambert}, {Lanza}, {Lasne}, {Le Campion}, {Le Fustec}, {Lebreton}, {Lebzelter}, {Leccia}, {Leclerc}, {Lecoeur-Taibi}, {Liao}, {Licata}, {Lindstr{\o}m}, {Lister}, {Livanou}, {Lobel}, {Madrero Pardo}, {Managau}, {Mann}, {Marchant}, {Marconi}, {Marcos Santos}, {Marinoni}, {Marocco}, {Marshall}, {Martin Polo}, {Mart{\'\i}n-Fleitas}, {Masip}, {Massari}, {Mastrobuono-Battisti}, {Mazeh}, {McMillan}, {Messina}, {Michalik}, {Millar}, {Mints}, {Molina}, {Molinaro}, {Moln{\'a}r}, {Montegriffo}, {Mor}, {Morbidelli}, {Morel}, {Morris}, {Mulone}, {Munoz}, {Muraveva}, {Murphy}, {Musella}, {Noval}, {Ord{\'e}novic}, {Orr{\`u}}, {Osinde}, {Pagani}, {Pagano},
  {Palaversa}, {Palicio}, {Panahi}, {Pawlak}, {Pe{\~n}alosa Esteller}, {Penttil{\"a}}, {Piersimoni}, {Pineau}, {Plachy}, {Plum}, {Poggio}, {Poretti}, {Poujoulet}, {Pr{\v{s}}a}, {Pulone}, {Racero}, {Ragaini}, {Rainer}, {Raiteri}, {Rambaux}, {Ramos}, {Ramos-Lerate}, {Re Fiorentin}, {Regibo}, {Reyl{\'e}}, {Ripepi}, {Riva}, {Rixon}, {Robichon}, {Robin}, {Roelens}, {Rohrbasser}, {Romero-G{\'o}mez}, {Rowell}, {Royer}, {Rybicki}, {Sadowski}, {Sagrist{\`a} Sell{\'e}s}, {Sahlmann}, {Salgado}, {Salguero}, {Samaras}, {Sanchez Gimenez}, {Sanna}, {Santove{\~n}a}, {Sarasso}, {Schultheis}, {Sciacca}, {Segol}, {Segovia}, {S{\'e}gransan}, {Semeux}, {Shahaf}, {Siddiqui}, {Siebert}, {Siltala}, {Slezak}, {Smart}, {Solano}, {Solitro}, {Souami}, {Souchay}, {Spagna}, {Spoto}, {Steele}, {Steidelm{\"u}ller}, {Stephenson}, {S{\"u}veges}, {Szabados}, {Szegedi-Elek}, {Taris}, {Tauran}, {Taylor}, {Teixeira}, {Thuillot}, {Tonello}, {Torra}, {Torra}, {Turon}, {Unger}, {Vaillant}, {van Dillen}, {Vanel}, {Vecchiato}, {Viala}, {Vicente},
  {Voutsinas}, {Weiler}, {Wevers}, {Wyrzykowski}, {Yoldas}, {Yvard}, {Zhao}, {Zorec}, {Zucker}, {Zurbach}, \& {Zwitter}}]{GaiaEDR3summary}
{Gaia Collaboration}, {Brown}, A.~G.~A., {Vallenari}, A., {et~al.} 2021, \aap, 649, A1, \dodoi{10.1051/0004-6361/202039657}

\bibitem[{{Gaia Collaboration} {et~al.}(2022){Gaia Collaboration}, {Arenou}, {Babusiaux}, {Barstow}, {Faigler}, {Jorissen}, {Kervella}, {Mazeh}, {Mowlavi}, {Panuzzo}, {Sahlmann}, {Shahaf}, {Sozzetti}, {Bauchet}, {Damerdji}, {Gavras}, {Giacobbe}, {Gosset}, {Halbwachs}, {Holl}, {Lattanzi}, {Leclerc}, {Morel}, {Pourbaix}, {Re Fiorentin}, {Sadowski}, {S{\'e}gransan}, {Siopis}, {Teyssier}, {Zwitter}, {Planquart}, {Brown}, {Vallenari}, {Prusti}, {de Bruijne}, {Biermann}, {Creevey}, {Ducourant}, {Evans}, {Eyer}, {Guerra}, {Hutton}, {Jordi}, {Klioner}, {Lammers}, {Lindegren}, {Luri}, {Mignard}, {Panem}, {Randich}, {Sartoretti}, {Soubiran}, {Tanga}, {Walton}, {Bailer-Jones}, {Bastian}, {Drimmel}, {Jansen}, {Katz}, {van Leeuwen}, {Bakker}, {Cacciari}, {Casta{\~n}eda}, {De Angeli}, {Fabricius}, {Fouesneau}, {Fr{\'e}mat}, {Galluccio}, {Guerrier}, {Heiter}, {Masana}, {Messineo}, {Nicolas}, {Nienartowicz}, {Pailler}, {Riclet}, {Roux}, {Seabroke}, {Sordo}, {Th{\'e}venin}, {Gracia-Abril}, {Portell}, {Altmann}, {Andrae},
  {Audard}, {Bellas-Velidis}, {Benson}, {Berthier}, {Blomme}, {Burgess}, {Busonero}, {Busso}, {C{\'a}novas}, {Carry}, {Cellino}, {Cheek}, {Clementini}, {Davidson}, {de Teodoro}, {Nu{\~n}ez Campos}, {Delchambre}, {Dell'Oro}, {Esquej}, {Fern{\'a}ndez-Hern{\'a}ndez}, {Fraile}, {Garabato}, {Garc{\'\i}a-Lario}, {Haigron}, {Hambly}, {Harrison}, {Hern{\'a}ndez}, {Hestroffer}, {Hodgkin}, {Jan{\ss}en}, {Jevardat de Fombelle}, {Jordan}, {Krone-Martins}, {Lanzafame}, {L{\"o}ffler}, {Marchal}, {Marrese}, {Moitinho}, {Muinonen}, {Osborne}, {Pancino}, {Pauwels}, {Recio-Blanco}, {Reyl{\'e}}, {Riello}, {Rimoldini}, {Roegiers}, {Rybizki}, {Sarro}, {Smith}, {Utrilla}, {van Leeuwen}, {Abbas}, {{\'A}brah{\'a}m}, {Abreu Aramburu}, {Aerts}, {Aguado}, {Ajaj}, {Aldea-Montero}, {Altavilla}, {{\'A}lvarez}, {Alves}, {Anders}, {Anderson}, {Anglada Varela}, {Antoja}, {Baines}, {Baker}, {Balaguer-N{\'u}{\~n}ez}, {Balbinot}, {Balog}, {Barache}, {Barbato}, {Barros}, {Bartolom{\'e}}, {Bassilana}, {Becciani}, {Bellazzini}, {Berihuete},
  {Bernet}, {Bertone}, {Bianchi}, {Binnenfeld}, {Blanco-Cuaresma}, {Blazere}, {Boch}, {Bombrun}, {Bossini}, {Bouquillon}, {Bragaglia}, {Bramante}, {Breedt}, {Bressan}, {Brouillet}, {Brugaletta}, {Bucciarelli}, {Burlacu}, {Butkevich}, {Buzzi}, {Caffau}, {Cancelliere}, {Cantat-Gaudin}, {Carballo}, {Carlucci}, {Carnerero}, {Carrasco}, {Casamiquela}, {Castellani}, {Castro-Ginard}, {Chaoul}, {Charlot}, {Chemin}, {Chiaramida}, {Chiavassa}, {Chornay}, {Comoretto}, {Contursi}, {Cooper}, {Cornez}, {Cowell}, {Crifo}, {Cropper}, {Crosta}, {Crowley}, {Dafonte}, {Dapergolas}, {David}, {de Laverny}, {De Luise}, {De March}, {De Ridder}, {de Souza}, {de Torres}, {del Peloso}, {del Pozo}, {Delbo}, {Delgado}, {Delisle}, {Demouchy}, {Dharmawardena}, {Diakite}, {Diener}, {Distefano}, {Dolding}, {Enke}, {Fabre}, {Fabrizio}, {Fedorets}, {Fernique}, {Figueras}, {Fournier}, {Fouron}, {Fragkoudi}, {Gai}, {Garcia-Gutierrez}, {Garcia-Reinaldos}, {Garc{\'\i}a-Torres}, {Garofalo}, {Gavel}, {Gerlach}, {Geyer}, {Gilmore}, {Girona},
  {Giuffrida}, {Gomel}, {Gomez}, {Gonz{\'a}lez-N{\'u}{\~n}ez}, {Gonz{\'a}lez-Santamar{\'\i}a}, {Gonz{\'a}lez-Vidal}, {Granvik}, {Guillout}, {Guiraud}, {Guti{\'e}rrez-S{\'a}nchez}, {Guy}, {Hatzidimitriou}, {Hauser}, {Haywood}, {Helmer}, {Helmi}, {Sarmiento}, {Hidalgo}, {H{\l}adczuk}, {Hobbs}, {Holland}, {Huckle}, {Jardine}, {Jasniewicz}, {Jean-Antoine Piccolo}, {Jim{\'e}nez-Arranz}, {Juaristi Campillo}, {Julbe}, {Karbevska}, {Khanna}, {Kordopatis}, {Korn}, {K{\'o}sp{\'a}l}, {Kostrzewa-Rutkowska}, {Kruszy{\'n}ska}, {Kun}, {Laizeau}, {Lambert}, {Lanza}, {Lasne}, {Le Campion}, {Lebreton}, {Lebzelter}, {Leccia}, {Lecoeur-Taibi}, {Liao}, {Licata}, {Lindstr{\o}m}, {Lister}, {Livanou}, {Lobel}, {Lorca}, {Loup}, {Madrero Pardo}, {Magdaleno Romeo}, {Managau}, {Mann}, {Manteiga}, {Marchant}, {Marconi}, {Marcos}, {Marcos Santos}, {Mar{\'\i}n Pina}, {Marinoni}, {Marocco}, {Marshall}, {Polo}, {Mart{\'\i}n-Fleitas}, {Marton}, {Mary}, {Masip}, {Massari}, {Mastrobuono-Battisti}, {McMillan}, {Messina}, {Michalik}, {Millar},
  {Mints}, {Molina}, {Molinaro}, {Moln{\'a}r}, {Monari}, {Mongui{\'o}}, {Montegriffo}, {Montero}, {Mor}, {Mora}, {Morbidelli}, {Morris}, {Muraveva}, {Murphy}, {Musella}, {Nagy}, {Noval}, {Oca{\~n}a}, {Ogden}, {Ordenovic}, {Osinde}, {Pagani}, {Pagano}, {Palaversa}, {Palicio}, {Pallas-Quintela}, {Panahi}, {Payne-Wardenaar}, {Pe{\~n}alosa Esteller}, {Penttil{\"a}}, {Pichon}, {Piersimoni}, {Pineau}, {Plachy}, {Plum}, {Poggio}, {Pr{\v{s}}a}, {Pulone}, {Racero}, {Ragaini}, {Rainer}, {Raiteri}, {Ramos}, {Ramos-Lerate}, {Regibo}, {Richards}, {Rios Diaz}, {Ripepi}, {Riva}, {Rix}, {Rixon}, {Robichon}, {Robin}, {Robin}, {Roelens}, {Rogues}, {Rohrbasser}, {Romero-G{\'o}mez}, {Rowell}, {Royer}, {Ruz Mieres}, {Rybicki}, {S{\'a}ez N{\'u}{\~n}ez}, {Sagrist{\`a} Sell{\'e}s}, {Salguero}, {Samaras}, {Sanchez Gimenez}, {Sanna}, {Santove{\~n}a}, {Sarasso}, {Schultheis}, {Sciacca}, {Segol}, {Segovia}, {Semeux}, {Siddiqui}, {Siebert}, {Siltala}, {Silvelo}, {Slezak}, {Slezak}, {Smart}, {Snaith}, {Solano}, {Solitro}, {Souami},
  {Souchay}, {Spagna}, {Spina}, {Spoto}, {Steele}, {Steidelm{\"u}ller}, {Stephenson}, {S{\"u}veges}, {Surdej}, {Szabados}, {Szegedi-Elek}, {Taris}, {Taylor}, {Teixeira}, {Tolomei}, {Tonello}, {Torra}, {Torra}, {Torralba Elipe}, {Trabucchi}, {Tsounis}, {Turon}, {Ulla}, {Unger}, {Vaillant}, {van Dillen}, {van Reeven}, {Vanel}, {Vecchiato}, {Viala}, {Vicente}, {Voutsinas}, {Weiler}, {Wevers}, {Wyrzykowski}, {Yoldas}, {Yvard}, {Zhao}, {Zorec}, \& {Zucker}}]{GaiaDR3binaries}
{Gaia Collaboration}, {Arenou}, F., {Babusiaux}, C., {et~al.} 2022, arXiv e-prints, arXiv:2206.05595.
\newblock \doarXiv{2206.05595}

\bibitem[{{Gaia Collaboration} {et~al.}(2023){Gaia Collaboration}, {Montegriffo}, {Bellazzini}, {De Angeli}, {Andrae}, {Barstow}, {Bossini}, {Bragaglia}, {Burgess}, {Cacciari}, {Carrasco}, {Chornay}, {Delchambre}, {Evans}, {Fouesneau}, {Fr{\'e}mat}, {Garabato}, {Jordi}, {Manteiga}, {Massari}, {Palaversa}, {Pancino}, {Riello}, {Ruz Mieres}, {Sanna}, {Santove{\~n}a}, {Sordo}, {Vallenari}, {Walton}, {Brown}, {Prusti}, {de Bruijne}, {Arenou}, {Babusiaux}, {Biermann}, {Creevey}, {Ducourant}, {Eyer}, {Guerra}, {Hutton}, {Klioner}, {Lammers}, {Lindegren}, {Luri}, {Mignard}, {Panem}, {Pourbaix}, {Randich}, {Sartoretti}, {Soubiran}, {Tanga}, {Bailer-Jones}, {Bastian}, {Drimmel}, {Jansen}, {Katz}, {Lattanzi}, {van Leeuwen}, {Bakker}, {Casta{\~n}eda}, {Fabricius}, {Galluccio}, {Guerrier}, {Heiter}, {Masana}, {Messineo}, {Mowlavi}, {Nicolas}, {Nienartowicz}, {Pailler}, {Panuzzo}, {Riclet}, {Roux}, {Seabroke}, {Th{\'e}venin}, {Gracia-Abril}, {Portell}, {Teyssier}, {Altmann}, {Audard}, {Bellas-Velidis}, {Benson},
  {Berthier}, {Blomme}, {Busonero}, {Busso}, {C{\'a}novas}, {Carry}, {Cellino}, {Cheek}, {Clementini}, {Damerdji}, {Davidson}, {de Teodoro}, {Nu{\~n}ez Campos}, {Dell'Oro}, {Esquej}, {Fern{\'a}ndez-Hern{\'a}ndez}, {Fraile}, {Garc{\'\i}a-Lario}, {Gosset}, {Haigron}, {Halbwachs}, {Hambly}, {Harrison}, {Hern{\'a}ndez}, {Hestroffer}, {Hodgkin}, {Holl}, {Jan{\ss}en}, {Jevardat de Fombelle}, {Jordan}, {Krone-Martins}, {Lanzafame}, {L{\"o}ffler}, {Marchal}, {Marrese}, {Moitinho}, {Muinonen}, {Osborne}, {Pauwels}, {Recio-Blanco}, {Reyl{\'e}}, {Rimoldini}, {Roegiers}, {Rybizki}, {Sarro}, {Siopis}, {Smith}, {Sozzetti}, {Utrilla}, {van Leeuwen}, {Abbas}, {{\'A}brah{\'a}m}, {Abreu Aramburu}, {Aerts}, {Aguado}, {Ajaj}, {Aldea-Montero}, {Altavilla}, {{\'A}lvarez}, {Alves}, {Anderson}, {Anglada Varela}, {Antoja}, {Baines}, {Baker}, {Balaguer-N{\'u}{\~n}ez}, {Balbinot}, {Balog}, {Barache}, {Barbato}, {Barros}, {Bartolom{\'e}}, {Bassilana}, {Bauchet}, {Becciani}, {Berihuete}, {Bernet}, {Bertone}, {Bianchi}, {Binnenfeld},
  {Blanco-Cuaresma}, {Boch}, {Bombrun}, {Bouquillon}, {Bramante}, {Breedt}, {Bressan}, {Brouillet}, {Brugaletta}, {Bucciarelli}, {Burlacu}, {Butkevich}, {Buzzi}, {Caffau}, {Cancelliere}, {Cantat-Gaudin}, {Carballo}, {Carlucci}, {Carnerero}, {Casamiquela}, {Castellani}, {Castro-Ginard}, {Chaoul}, {Charlot}, {Chemin}, {Chiaramida}, {Chiavassa}, {Comoretto}, {Contursi}, {Cooper}, {Cornez}, {Cowell}, {Crifo}, {Cropper}, {Crosta}, {Crowley}, {Dafonte}, {Dapergolas}, {David}, {de Laverny}, {De Luise}, {De March}, {De Ridder}, {de Souza}, {de Torres}, {del Peloso}, {del Pozo}, {Delbo}, {Delgado}, {Delisle}, {Demouchy}, {Dharmawardena}, {Diakite}, {Diener}, {Distefano}, {Dolding}, {Enke}, {Fabre}, {Fabrizio}, {Faigler}, {Fedorets}, {Fernique}, {Figueras}, {Fournier}, {Fouron}, {Fragkoudi}, {Gai}, {Garcia-Gutierrez}, {Garcia-Reinaldos}, {Garc{\'\i}a-Torres}, {Garofalo}, {Gavel}, {Gavras}, {Gerlach}, {Geyer}, {Giacobbe}, {Gilmore}, {Girona}, {Giuffrida}, {Gomel}, {Gomez}, {Gonz{\'a}lez-N{\'u}{\~n}ez},
  {Gonz{\'a}lez-Santamar{\'\i}a}, {Gonz{\'a}lez-Vidal}, {Granvik}, {Guillout}, {Guiraud}, {Guti{\'e}rrez-S{\'a}nchez}, {Guy}, {Hatzidimitriou}, {Hauser}, {Haywood}, {Helmer}, {Helmi}, {Sarmiento}, {Hidalgo}, {H{\l}adczuk}, {Hobbs}, {Holland}, {Huckle}, {Jardine}, {Jasniewicz}, {Jean-Antoine Piccolo}, {Jim{\'e}nez-Arranz}, {Juaristi Campillo}, {Julbe}, {Karbevska}, {Kervella}, {Khanna}, {Kordopatis}, {Korn}, {K{\'o}sp{\'a}l}, {Kostrzewa-Rutkowska}, {Kruszy{\'n}ska}, {Kun}, {Laizeau}, {Lambert}, {Lanza}, {Lasne}, {Le Campion}, {Lebreton}, {Lebzelter}, {Leccia}, {Leclerc}, {Lecoeur-Taibi}, {Liao}, {Licata}, {Lindstr{\'o}m}, {Lister}, {Livanou}, {Lobel}, {Lorca}, {Loup}, {Madrero Pardo}, {Magdaleno Romeo}, {Managau}, {Mann}, {Marchant}, {Marconi}, {Marcos}, {Marcos Santos}, {Mar{\'\i}n Pina}, {Marinoni}, {Marocco}, {Marshall}, {Martin Polo}, {Mart{\'\i}n-Fleitas}, {Marton}, {Mary}, {Masip}, {Mastrobuono-Battisti}, {Mazeh}, {McMillan}, {Messina}, {Michalik}, {Millar}, {Mints}, {Molina}, {Molinaro}, {Moln{\'a}r},
  {Monari}, {Mongui{\'o}}, {Montero}, {Mor}, {Mora}, {Morbidelli}, {Morel}, {Morris}, {Muraveva}, {Murphy}, {Musella}, {Nagy}, {Noval}, {Oca{\~n}a}, {Ogden}, {Ordenovic}, {Osinde}, {Pagani}, {Pagano}, {Palicio}, {Pallas-Quintela}, {Panahi}, {Payne-Wardenaar}, {Pe{\~n}alosa Esteller}, {Penttil{\"a}}, {Pichon}, {Piersimoni}, {Pineau}, {Plachy}, {Plum}, {Poggio}, {Pr{\v{s}}a}, {Pulone}, {Racero}, {Ragaini}, {Rainer}, {Raiteri}, {Ramos}, {Ramos-Lerate}, {Re Fiorentin}, {Regibo}, {Richards}, {Rios Diaz}, {Ripepi}, {Riva}, {Rix}, {Rixon}, {Robichon}, {Robin}, {Robin}, {Roelens}, {Rogues}, {Rohrbasser}, {Romero-G{\'o}mez}, {Rowell}, {Royer}, {Rybicki}, {Sadowski}, {S{\'a}ez N{\'u}{\~n}ez}, {Sagrist{\`a} Sell{\'e}s}, {Sahlmann}, {Salguero}, {Samaras}, {Sanchez Gimenez}, {Sarasso}, {Schultheis}, {Sciacca}, {Segol}, {Segovia}, {S{\'e}gransan}, {Semeux}, {Shahaf}, {Siddiqui}, {Siebert}, {Siltala}, {Silvelo}, {Slezak}, {Slezak}, {Smart}, {Snaith}, {Solano}, {Solitro}, {Souami}, {Souchay}, {Spagna}, {Spina}, {Spoto},
  {Steele}, {Steidelm{\"u}ller}, {Stephenson}, {S{\"u}veges}, {Surdej}, {Szabados}, {Szegedi-Elek}, {Taris}, {Taylor}, {Teixeira}, {Tolomei}, {Tonello}, {Torra}, {Torra}, {Torralba Elipe}, {Trabucchi}, {Tsounis}, {Turon}, {Ulla}, {Unger}, {Vaillant}, {van Dillen}, {van Reeven}, {Vanel}, {Vecchiato}, {Viala}, {Vicente}, {Voutsinas}, {Wevers}, {Wyrzykowski}, {Yoldas}, {Yvard}, {Zhao}, {Zorec}, {Zucker}, \& {Zwitter}}]{GaiaSynthPhot}
{Gaia Collaboration}, {Montegriffo}, P., {Bellazzini}, M., {et~al.} 2023, \aap, 674, A33, \dodoi{10.1051/0004-6361/202243709}

\bibitem[{{Gao} {et~al.}(2022){Gao}, {Toonen}, \& {Leigh}}]{Gao22}
{Gao}, Y., {Toonen}, S., \& {Leigh}, N. 2022, arXiv e-prints, arXiv:2203.05357.
\newblock \doarXiv{2203.05357}

\bibitem[{{Gontcharov}(2012)}]{Gontcharov2012}
{Gontcharov}, G.~A. 2012, Astronomy Letters, 38, 87, \dodoi{10.1134/S1063773712010033}

\bibitem[{{Goswami} {et~al.}(2023){Goswami}, {Shejeelammal}, {Goswami}, \& {Purandardas}}]{2023Goswami}
{Goswami}, A., {Shejeelammal}, J., {Goswami}, P.~P., \& {Purandardas}, M. 2023, arXiv e-prints, arXiv:2311.10043.
\newblock \doarXiv{2311.10043}

\bibitem[{{Gray} {et~al.}(2011){Gray}, {McGahee}, {Griffin}, \& {Corbally}}]{Gray11}
{Gray}, R.~O., {McGahee}, C.~E., {Griffin}, R.~E.~M., \& {Corbally}, C.~J. 2011, AJ, 141, 160, \dodoi{10.1088/0004-6256/141/5/160}

\bibitem[{{Gustafsson} {et~al.}(2008){Gustafsson}, {Edvardsson}, {Eriksson}, {J{\o}rgensen}, {Nordlund}, \& {Plez}}]{2008A&A...486..951G}
{Gustafsson}, B., {Edvardsson}, B., {Eriksson}, K., {et~al.} 2008, \aap, 486, 951, \dodoi{10.1051/0004-6361:200809724}

\bibitem[{{Hansen} {et~al.}(2012){Hansen}, {Bergemann}, {Cescutti}, {Francois}, {Arcones}, {Karakas}, {Lind}, \& {Chiappini}}]{2012Hansen}
{Hansen}, C.~J., {Bergemann}, M., {Cescutti}, G., {et~al.} 2012, arXiv e-prints, arXiv:1212.4147, \dodoi{10.48550/arXiv.1212.4147}

\bibitem[{{Heiter} {et~al.}(2021){Heiter}, {Lind}, {Bergemann}, {Asplund}, {Mikolaitis}, {Barklem}, {Masseron}, {de Laverny}, {Magrini}, {Edvardsson}, {J{\"o}nsson}, {Pickering}, {Ryde}, {Bayo Ar{\'a}n}, {Bensby}, {Casey}, {Feltzing}, {Jofr{\'e}}, {Korn}, {Pancino}, {Damiani}, {Lanzafame}, {Lardo}, {Monaco}, {Morbidelli}, {Smiljanic}, {Worley}, {Zaggia}, {Randich}, \& {Gilmore}}]{2021Heiter}
{Heiter}, U., {Lind}, K., {Bergemann}, M., {et~al.} 2021, \aap, 645, A106, \dodoi{10.1051/0004-6361/201936291}

\bibitem[{{Henden} {et~al.}(2015){Henden}, {Levine}, {Terrell}, \& {Welch}}]{Henden2015}
{Henden}, A.~A., {Levine}, S., {Terrell}, D., \& {Welch}, D.~L. 2015, in American Astronomical Society Meeting Abstracts, Vol. 225, American Astronomical Society Meeting Abstracts \#225, 336.16

\bibitem[{{H{\o}g} {et~al.}(2000){H{\o}g}, {Fabricius}, {Makarov}, {Urban}, {Corbin}, {Wycoff}, {Bastian}, {Schwekendiek}, \& {Wicenec}}]{Hog2000}
{H{\o}g}, E., {Fabricius}, C., {Makarov}, V.~V., {et~al.} 2000, \aap, 355, L27

\bibitem[{{Howell} {et~al.}(2014){Howell}, {Sobeck}, {Haas}, {Still}, {Barclay}, {Mullally}, {Troeltzsch}, {Aigrain}, {Bryson}, {Caldwell}, {Chaplin}, {Cochran}, {Huber}, {Marcy}, {Miglio}, {Najita}, {Smith}, {Twicken}, \& {Fortney}}]{2014PASP..126..398H}
{Howell}, S.~B., {Sobeck}, C., {Haas}, M., {et~al.} 2014, \pasp, 126, 398, \dodoi{10.1086/676406}

\bibitem[{{Huber} {et~al.}(2009){Huber}, {Stello}, {Bedding}, {Chaplin}, {Arentoft}, {Quirion}, \& {Kjeldsen}}]{Huber2009}
{Huber}, D., {Stello}, D., {Bedding}, T.~R., {et~al.} 2009, Communications in Asteroseismology, 160, 74, \dodoi{10.48550/arXiv.0910.2764}

\bibitem[{{Husti} {et~al.}(2009){Husti}, {Gallino}, {Bisterzo}, {Straniero}, \& {Cristallo}}]{Husti09}
{Husti}, L., {Gallino}, R., {Bisterzo}, S., {Straniero}, O., \& {Cristallo}, S. 2009, \pasa, 26, 176, \dodoi{10.1071/AS08065}

\bibitem[{{Indebetouw} {et~al.}(2005){Indebetouw}, {Mathis}, {Babler}, {Meade}, {Watson}, {Whitney}, {Wolff}, {Wolfire}, {Cohen}, {Bania}, {Benjamin}, {Clemens}, {Dickey}, {Jackson}, {Kobulnicky}, {Marston}, {Mercer}, {Stauffer}, {Stolovy}, \& {Churchwell}}]{Indebetouw05}
{Indebetouw}, R., {Mathis}, J.~S., {Babler}, B.~L., {et~al.} 2005, \apj, 619, 931, \dodoi{10.1086/426679}

\bibitem[{{Izzard} {et~al.}(2010){Izzard}, {Dermine}, \& {Church}}]{Izzard10}
{Izzard}, R.~G., {Dermine}, T., \& {Church}, R.~P. 2010, \aap, 523, A10, \dodoi{10.1051/0004-6361/201015254}

\bibitem[{{Jonsell} {et~al.}(2006){Jonsell}, {Barklem}, {Gustafsson}, {Christlieb}, {Hill}, {Beers}, \& {Holmberg}}]{2006Jonsell}
{Jonsell}, K., {Barklem}, P.~S., {Gustafsson}, B., {et~al.} 2006, \aap, 451, 651, \dodoi{10.1051/0004-6361:20054470}

\bibitem[{{Jorissen} {et~al.}(2019){Jorissen}, {Boffin}, {Karinkuzhi}, {Van Eck}, {Escorza}, {Shetye}, \& {Van Winckel}}]{Jorissen19}
{Jorissen}, A., {Boffin}, H.~M.~J., {Karinkuzhi}, D., {et~al.} 2019, \aap, 626, A127, \dodoi{10.1051/0004-6361/201834630}

\bibitem[{{Jorissen} {et~al.}(1998){Jorissen}, {Van Eck}, {Mayor}, \& {Udry}}]{Jorissen98}
{Jorissen}, A., {Van Eck}, S., {Mayor}, M., \& {Udry}, S. 1998, \aap, 332, 877

\bibitem[{{K{\"a}ppeler} {et~al.}(2011){K{\"a}ppeler}, {Gallino}, {Bisterzo}, \& {Aoki}}]{Kappeler11}
{K{\"a}ppeler}, F., {Gallino}, R., {Bisterzo}, S., \& {Aoki}, W. 2011, Reviews of Modern Physics, 83, 157, \dodoi{10.1103/RevModPhys.83.157}

\bibitem[{{Karakas}(2010)}]{Karakas10}
{Karakas}, A.~I. 2010, \mnras, 403, 1413, \dodoi{10.1111/j.1365-2966.2009.16198.x}

\bibitem[{{Karakas} \& {Lattanzio}(2003)}]{2003Karakas}
{Karakas}, A.~I., \& {Lattanzio}, J.~C. 2003, \pasa, 20, 279, \dodoi{10.1071/AS03010}

\bibitem[{{Karakas} \& {Lattanzio}(2014)}]{KarakasLattanzio14}
---. 2014, \pasa, 31, e030, \dodoi{10.1017/pasa.2014.21}

\bibitem[{{Karinkuzhi} {et~al.}(2021){Karinkuzhi}, {Van Eck}, {Goriely}, {Siess}, {Jorissen}, {Merle}, {Escorza}, \& {Masseron}}]{Karinkuzhi21}
{Karinkuzhi}, D., {Van Eck}, S., {Goriely}, S., {et~al.} 2021, \aap, 645, A61, \dodoi{10.1051/0004-6361/202038891}

\bibitem[{{Karinkuzhi} {et~al.}(2018){Karinkuzhi}, {Van Eck}, {Jorissen}, {Goriely}, {Siess}, {Merle}, {Escorza}, {Van der Swaelmen}, {Boffin}, {Masseron}, {Shetye}, \& {Plez}}]{Karinkuzhi18}
{Karinkuzhi}, D., {Van Eck}, S., {Jorissen}, A., {et~al.} 2018, \aap, 618, A32, \dodoi{10.1051/0004-6361/201833084}

\bibitem[{{Karinkuzhi} {et~al.}(2023){Karinkuzhi}, {Van Eck}, {Goriely}, {Siess}, {Jorissen}, {Choplin}, {Escorza}, {Shetye}, \& {Van Winckel}}]{Karinkuzhi23}
{Karinkuzhi}, D., {Van Eck}, S., {Goriely}, S., {et~al.} 2023, \aap, 677, A47, \dodoi{10.1051/0004-6361/202345991}

\bibitem[{{Keenan}(1942)}]{Keenan42}
{Keenan}, P.~C. 1942, \apj, 96, 101, \dodoi{10.1086/144435}

\bibitem[{{Kervella} {et~al.}(2022){Kervella}, {Arenou}, \& {Th{\'e}venin}}]{Kervella22}
{Kervella}, P., {Arenou}, F., \& {Th{\'e}venin}, F. 2022, \aap, 657, A7, \dodoi{10.1051/0004-6361/202142146}

\bibitem[{{Kjeldsen} \& {Bedding}(1995)}]{1995Kjeldsen}
{Kjeldsen}, H., \& {Bedding}, T.~R. 1995, \aap, 293, 87, \dodoi{10.48550/arXiv.astro-ph/9403015}

\bibitem[{{Korotin} {et~al.}(2015){Korotin}, {Andrievsky}, {Hansen}, {Caffau}, {Bonifacio}, {Spite}, {Spite}, \& {Fran{\c{c}}ois}}]{2015Korotin}
{Korotin}, S.~A., {Andrievsky}, S.~M., {Hansen}, C.~J., {et~al.} 2015, \aap, 581, A70, \dodoi{10.1051/0004-6361/201526558}

\bibitem[{Kos {et~al.}(2016)Kos, Lin, Zwitter, Žerjal, Sharma, Bland-Hawthorn, Asplund, Casey, De~Silva, Freeman, Martell, Simpson, Schlesinger, Zucker, Anguiano, Bacigalupo, Bedding, Betters, Da~Costa, Duong, Hyde, Ireland, Kafle, Leon-Saval, Lewis, Munari, Nataf, Stello, Tinney, Traven, Watson, \& Wittenmyer}]{Kos2016}
Kos, J., Lin, J., Zwitter, T., {et~al.} 2016, Monthly Notices of the Royal Astronomical Society, 464, 1259, \dodoi{10.1093/mnras/stw2064}

\bibitem[{{Lawler} {et~al.}(2001){Lawler}, {Bonvallet}, \& {Sneden}}]{Lawler2001}
{Lawler}, J.~E., {Bonvallet}, G., \& {Sneden}, C. 2001, \apj, 556, 452, \dodoi{10.1086/321549}

\bibitem[{{Li} {et~al.}(2023){Li}, {Huang}, {Dong}, {Chen}, \& {Luo}}]{APOGEE-RVstds}
{Li}, Q.-Z., {Huang}, Y., {Dong}, X.-B., {Chen}, J.-J., \& {Luo}, A.~L. 2023, Research in Astronomy and Astrophysics, 23, 115026, \dodoi{10.1088/1674-4527/acf1e5}

\bibitem[{{Lindegren} {et~al.}(2018){Lindegren}, {Hern{\'a}ndez}, {Bombrun}, {Klioner}, {Bastian}, {Ramos-Lerate}, {de Torres}, {Steidelm{\"u}ller}, {Stephenson}, {Hobbs}, {Lammers}, {Biermann}, {Geyer}, {Hilger}, {Michalik}, {Stampa}, {McMillan}, {Casta{\~n}eda}, {Clotet}, {Comoretto}, {Davidson}, {Fabricius}, {Gracia}, {Hambly}, {Hutton}, {Mora}, {Portell}, {van Leeuwen}, {Abbas}, {Abreu}, {Altmann}, {Andrei}, {Anglada}, {Balaguer-N{\'u}{\~n}ez}, {Barache}, {Becciani}, {Bertone}, {Bianchi}, {Bouquillon}, {Bourda}, {Br{\"u}semeister}, {Bucciarelli}, {Busonero}, {Buzzi}, {Cancelliere}, {Carlucci}, {Charlot}, {Cheek}, {Crosta}, {Crowley}, {de Bruijne}, {de Felice}, {Drimmel}, {Esquej}, {Fienga}, {Fraile}, {Gai}, {Garralda}, {Gonz{\'a}lez-Vidal}, {Guerra}, {Hauser}, {Hofmann}, {Holl}, {Jordan}, {Lattanzi}, {Lenhardt}, {Liao}, {Licata}, {Lister}, {L{\"o}ffler}, {Marchant}, {Martin-Fleitas}, {Messineo}, {Mignard}, {Morbidelli}, {Poggio}, {Riva}, {Rowell}, {Salguero}, {Sarasso}, {Sciacca}, {Siddiqui}, {Smart},
  {Spagna}, {Steele}, {Taris}, {Torra}, {van Elteren}, {van Reeven}, \& {Vecchiato}}]{Lindegren18}
{Lindegren}, L., {Hern{\'a}ndez}, J., {Bombrun}, A., {et~al.} 2018, \aap, 616, A2, \dodoi{10.1051/0004-6361/201832727}

\bibitem[{{Liu} {et~al.}(2020){Liu}, {Shi}, \& {Wu}}]{2020Liu}
{Liu}, S., {Shi}, J., \& {Wu}, Z. 2020, \apj, 896, 64, \dodoi{10.3847/1538-4357/ab8f33}

\bibitem[{{Longland} {et~al.}(2012){Longland}, {Iliadis}, \& {Karakas}}]{2012Longland}
{Longland}, R., {Iliadis}, C., \& {Karakas}, A.~I. 2012, \prc, 85, 065809, \dodoi{10.1103/PhysRevC.85.065809}

\bibitem[{{Lugaro} {et~al.}(2003{\natexlab{a}}){Lugaro}, {Davis}, {Gallino}, {Pellin}, {Straniero}, \& {K{\"a}ppeler}}]{Lugaro03}
{Lugaro}, M., {Davis}, A.~M., {Gallino}, R., {et~al.} 2003{\natexlab{a}}, \apj, 593, 486, \dodoi{10.1086/376442}

\bibitem[{{Lugaro} {et~al.}(2003{\natexlab{b}}){Lugaro}, {Herwig}, {Lattanzio}, {Gallino}, \& {Straniero}}]{Lugaro03long}
{Lugaro}, M., {Herwig}, F., {Lattanzio}, J.~C., {Gallino}, R., \& {Straniero}, O. 2003{\natexlab{b}}, \apj, 586, 1305, \dodoi{10.1086/367887}

\bibitem[{{Lugaro} {et~al.}(2003{\natexlab{c}}){Lugaro}, {Herwig}, {Lattanzio}, {Gallino}, \& {Straniero}}]{2003Lugaro}
---. 2003{\natexlab{c}}, \apj, 586, 1305, \dodoi{10.1086/367887}

\bibitem[{{Lugaro} {et~al.}(2012{\natexlab{a}}){Lugaro}, {Karakas}, {Stancliffe}, \& {Rijs}}]{Lugaro12}
{Lugaro}, M., {Karakas}, A.~I., {Stancliffe}, R.~J., \& {Rijs}, C. 2012{\natexlab{a}}, \apj, 747, 2, \dodoi{10.1088/0004-637X/747/1/2}

\bibitem[{{Lugaro} {et~al.}(2012{\natexlab{b}}){Lugaro}, {Karakas}, {Stancliffe}, \& {Rijs}}]{2012Lugaro}
---. 2012{\natexlab{b}}, \apj, 747, 2, \dodoi{10.1088/0004-637X/747/1/2}

\bibitem[{{Lugaro} {et~al.}(2023){Lugaro}, {Pignatari}, {Reifarth}, \& {Wiescher}}]{2023Lugarorew}
{Lugaro}, M., {Pignatari}, M., {Reifarth}, R., \& {Wiescher}, M. 2023, Annual Review of Nuclear and Particle Science, 73, 315, \dodoi{10.1146/annurev-nucl-102422-080857}

\bibitem[{{Lugaro} {et~al.}(2016){Lugaro}, {Campbell}, {D'Orazi}, {Karakas}, {Garcia-Hernandez}, {Stancliffe}, {Tagliente}, {Iliadis}, \& {Rauscher}}]{Lugaro16}
{Lugaro}, M., {Campbell}, S.~W., {D'Orazi}, V., {et~al.} 2016, in Journal of Physics Conference Series, Vol. 665, 012021, \dodoi{10.1088/1742-6596/665/1/012021}

\bibitem[{{Maiorca} {et~al.}(2011){Maiorca}, {Randich}, {Busso}, {Magrini}, \& {Palmerini}}]{Maiorca2011}
{Maiorca}, E., {Randich}, S., {Busso}, M., {Magrini}, L., \& {Palmerini}, S. 2011, in Astronomical Society of the Pacific Conference Series, Vol. 445, Why Galaxies Care about AGB Stars II: Shining Examples and Common Inhabitants, ed. F.~{Kerschbaum}, T.~{Lebzelter}, \& R.~F. {Wing}, 159

\bibitem[{{Majewski} {et~al.}(2017){Majewski}, {Schiavon}, {Frinchaboy}, {Allende Prieto}, {Barkhouser}, {Bizyaev}, {Blank}, {Brunner}, {Burton}, {Carrera}, {Chojnowski}, {Cunha}, {Epstein}, {Fitzgerald}, {Garc{\'\i}a P{\'e}rez}, {Hearty}, {Henderson}, {Holtzman}, {Johnson}, {Lam}, {Lawler}, {Maseman}, {M{\'e}sz{\'a}ros}, {Nelson}, {Nguyen}, {Nidever}, {Pinsonneault}, {Shetrone}, {Smee}, {Smith}, {Stolberg}, {Skrutskie}, {Walker}, {Wilson}, {Zasowski}, {Anders}, {Basu}, {Beland}, {Blanton}, {Bovy}, {Brownstein}, {Carlberg}, {Chaplin}, {Chiappini}, {Eisenstein}, {Elsworth}, {Feuillet}, {Fleming}, {Galbraith-Frew}, {Garc{\'\i}a}, {Garc{\'\i}a-Hern{\'a}ndez}, {Gillespie}, {Girardi}, {Gunn}, {Hasselquist}, {Hayden}, {Hekker}, {Ivans}, {Kinemuchi}, {Klaene}, {Mahadevan}, {Mathur}, {Mosser}, {Muna}, {Munn}, {Nichol}, {O'Connell}, {Parejko}, {Robin}, {Rocha-Pinto}, {Schultheis}, {Serenelli}, {Shane}, {Silva Aguirre}, {Sobeck}, {Thompson}, {Troup}, {Weinberg}, \& {Zamora}}]{APOGEEsurvey}
{Majewski}, S.~R., {Schiavon}, R.~P., {Frinchaboy}, P.~M., {et~al.} 2017, \aj, 154, 94, \dodoi{10.3847/1538-3881/aa784d}

\bibitem[{{Mashonkina} \& {Christlieb}(2014)}]{2014Mashonkina}
{Mashonkina}, L., \& {Christlieb}, N. 2014, \aap, 565, A123, \dodoi{10.1051/0004-6361/201423651}

\bibitem[{{Mashonkina} {et~al.}(2003){Mashonkina}, {Gehren}, {Travaglio}, \& {Borkova}}]{2003Mashonkina}
{Mashonkina}, L., {Gehren}, T., {Travaglio}, C., \& {Borkova}, T. 2003, \aap, 397, 275, \dodoi{10.1051/0004-6361:20021512}

\bibitem[{{Mashonkina} {et~al.}(2017){Mashonkina}, {Jablonka}, {Sitnova}, {Pakhomov}, \& {North}}]{2017Mashonkina}
{Mashonkina}, L., {Jablonka}, P., {Sitnova}, T., {Pakhomov}, Y., \& {North}, P. 2017, \aap, 608, A89, \dodoi{10.1051/0004-6361/201731582}

\bibitem[{{Masseron} {et~al.}(2010){Masseron}, {Johnson}, {Plez}, {van Eck}, {Primas}, {Goriely}, \& {Jorissen}}]{Masseron2010}
{Masseron}, T., {Johnson}, J.~A., {Plez}, B., {et~al.} 2010, \aap, 509, A93, \dodoi{10.1051/0004-6361/200911744}

\bibitem[{{McClure}(1984)}]{McClure84}
{McClure}, R.~D. 1984, PASP, 96, 117, \dodoi{10.1086/131310}

\bibitem[{{Mennessier} {et~al.}(1997){Mennessier}, {Luri}, {Figueras}, {Gomez}, {Grenier}, {Torra}, \& {North}}]{Mennessier97}
{Mennessier}, M.~O., {Luri}, X., {Figueras}, F., {et~al.} 1997, \aap, 326, 722

\bibitem[{{Mohamed} \& {Podsiadlowski}(2007)}]{Mohamed07}
{Mohamed}, S., \& {Podsiadlowski}, P. 2007, in Astronomical Society of the Pacific Conference Series, Vol. 372, 15th European Workshop on White Dwarfs, ed. R.~{Napiwotzki} \& M.~R. {Burleigh}, 397

\bibitem[{{Neyskens} {et~al.}(2015){Neyskens}, {van Eck}, {Jorissen}, {Goriely}, {Siess}, \& {Plez}}]{2015Neyskens}
{Neyskens}, P., {van Eck}, S., {Jorissen}, A., {et~al.} 2015, \nat, 517, 174, \dodoi{10.1038/nature14050}

\bibitem[{{Nidever}(2021)}]{Nidever-doppler}
{Nidever}, D. 2021, Zenodo, \dodoi{10.5281/zenodo.4906681}

\bibitem[{{North} {et~al.}(2000){North}, {Jorissen}, \& {Mayor}}]{North00}
{North}, P., {Jorissen}, A., \& {Mayor}, M. 2000, in IAU Symposium, Vol. 177, The Carbon Star Phenomenon, ed. R.~F. {Wing}, 269

\bibitem[{Ochsenbein(1996)}]{10.26093/cds/vizier}
Ochsenbein, F. 1996, The VizieR database of astronomical catalogues,  CDS, Centre de DonnÃ©es astronomiques de Strasbourg, \dodoi{10.26093/CDS/VIZIER}

\bibitem[{{Ochsenbein} {et~al.}(2000){Ochsenbein}, {Bauer}, \& {Marcout}}]{vizier2000}
{Ochsenbein}, F., {Bauer}, P., \& {Marcout}, J. 2000, \aaps, 143, 23, \dodoi{10.1051/aas:2000169}

\bibitem[{{Pepe} {et~al.}(2002){Pepe}, {Mayor}, {Galland}, {Naef}, {Queloz}, {Santos}, {Udry}, \& {Burnet}}]{CCF}
{Pepe}, F., {Mayor}, M., {Galland}, F., {et~al.} 2002, \aap, 388, 632, \dodoi{10.1051/0004-6361:20020433}

\bibitem[{{Pignatari} {et~al.}(2010){Pignatari}, {Gallino}, {Heil}, {Wiescher}, {K{\"a}ppeler}, {Herwig}, \& {Bisterzo}}]{2010ApJ...710.1557P}
{Pignatari}, M., {Gallino}, R., {Heil}, M., {et~al.} 2010, \apj, 710, 1557, \dodoi{10.1088/0004-637X/710/2/1557}

\bibitem[{{Plez}(2012)}]{2012ascl.soft05004P}
{Plez}, B. 2012, {Turbospectrum: Code for spectral synthesis}, Astrophysics Source Code Library, record ascl:1205.004.
\newblock \doeprint{1205.004}

\bibitem[{{Pols} {et~al.}(2003){Pols}, {Karakas}, {Lattanzio}, \& {Tout}}]{Pols03}
{Pols}, O.~R., {Karakas}, A.~I., {Lattanzio}, J.~C., \& {Tout}, C.~A. 2003, in Astronomical Society of the Pacific Conference Series, Vol. 303, Symbiotic Stars Probing Stellar Evolution, ed. R.~L.~M. {Corradi}, J.~{Mikolajewska}, \& T.~J. {Mahoney}, 290

\bibitem[{{Pourbaix} \& {Jorissen}(2000)}]{Pourbaix2000}
{Pourbaix}, D., \& {Jorissen}, A. 2000, \aaps, 145, 161, \dodoi{10.1051/aas:2000346}

\bibitem[{{Prantzos} {et~al.}(2020){Prantzos}, {Abia}, {Cristallo}, {Limongi}, \& {Chieffi}}]{2020Prantzos}
{Prantzos}, N., {Abia}, C., {Cristallo}, S., {Limongi}, M., \& {Chieffi}, A. 2020, \mnras, 491, 1832, \dodoi{10.1093/mnras/stz3154}

\bibitem[{{Ram{\'\i}rez} \& {Allende Prieto}(2011)}]{Arcturus}
{Ram{\'\i}rez}, I., \& {Allende Prieto}, C. 2011, \apj, 743, 135, \dodoi{10.1088/0004-637X/743/2/135}

\bibitem[{{Roederer} {et~al.}(2016){Roederer}, {Karakas}, {Pignatari}, \& {Herwig}}]{Roederer16}
{Roederer}, I.~U., {Karakas}, A.~I., {Pignatari}, M., \& {Herwig}, F. 2016, \apj, 821, 37, \dodoi{10.3847/0004-637X/821/1/37}

\bibitem[{{Roriz} {et~al.}(2024){Roriz}, {Lugaro}, {Junqueira}, {Sneden}, {Drake}, \& {Pereira}}]{2024RorizTungsten}
{Roriz}, M.~P., {Lugaro}, M., {Junqueira}, S., {et~al.} 2024, \mnras, 528, 4354, \dodoi{10.1093/mnras/stae221}

\bibitem[{{Roriz} {et~al.}(2021{\natexlab{a}}){Roriz}, {Lugaro}, {Pereira}, {Drake}, {Junqueira}, \& {Sneden}}]{2021RorizRb}
{Roriz}, M.~P., {Lugaro}, M., {Pereira}, C.~B., {et~al.} 2021{\natexlab{a}}, \mnras, 501, 5834, \dodoi{10.1093/mnras/staa3888}

\bibitem[{{Roriz} {et~al.}(2021{\natexlab{b}}){Roriz}, {Lugaro}, {Pereira}, {Sneden}, {Junqueira}, {Karakas}, \& {Drake}}]{Roriz21}
---. 2021{\natexlab{b}}, \mnras, 507, 1956, \dodoi{10.1093/mnras/stab2014}

\bibitem[{{Roriz} {et~al.}(2023){Roriz}, {Pereira}, {Junqueira}, {Lugaro}, {Drake}, \& {Sneden}}]{2023Roriz}
{Roriz}, M.~P., {Pereira}, C.~B., {Junqueira}, S., {et~al.} 2023, \mnras, 518, 5414, \dodoi{10.1093/mnras/stac3378}

\bibitem[{{Saladino} \& {Pols}(2019)}]{Saladino19}
{Saladino}, M.~I., \& {Pols}, O.~R. 2019, arXiv e-prints, arXiv:1906.02038.
\newblock \doarXiv{1906.02038}

\bibitem[{{Shaltout} {et~al.}(2020){Shaltout}, {Abdelkawy}, \& {Beheary}}]{2020Shaltout}
{Shaltout}, A. M.~K., {Abdelkawy}, A. G.~A., \& {Beheary}, M.~M. 2020, \mnras, 496, 5361, \dodoi{10.1093/mnras/staa1825}

\bibitem[{{Shejeelammal} {et~al.}(2020){Shejeelammal}, {Goswami}, {Goswami}, {Rathour}, \& {Masseron}}]{Shejeelammal20}
{Shejeelammal}, J., {Goswami}, A., {Goswami}, P.~P., {Rathour}, R.~S., \& {Masseron}, T. 2020, \mnras, 492, 3708, \dodoi{10.1093/mnras/stz3518}

\bibitem[{{Shetye} {et~al.}(2020){Shetye}, {Van Eck}, {Goriely}, {Siess}, {Jorissen}, {Escorza}, \& {Van Winckel}}]{Shetye20}
{Shetye}, S., {Van Eck}, S., {Goriely}, S., {et~al.} 2020, \aap, 635, L6, \dodoi{10.1051/0004-6361/202037481}

\bibitem[{{Shetye} {et~al.}(2018){Shetye}, {Van Eck}, {Jorissen}, {Van Winckel}, {Siess}, {Goriely}, {Escorza}, {Karinkuzhi}, \& {Plez}}]{Shetye18}
{Shetye}, S., {Van Eck}, S., {Jorissen}, A., {et~al.} 2018, \aap, 620, A148, \dodoi{10.1051/0004-6361/201833298}

\bibitem[{{Siess}(2008)}]{Siess08}
{Siess}, L. 2008, EAS Publications Series, 32, 131, \dodoi{10.1051/eas:0832004}

\bibitem[{{Siess} {et~al.}(2000){Siess}, {Dufour}, \& {Forestini}}]{Siess00}
{Siess}, L., {Dufour}, E., \& {Forestini}, M. 2000, \aap, 358, 593

\bibitem[{{Silverans} {et~al.}(1986){Silverans}, {Borghs}, {de Bisschop}, \& {van Hove}}]{1986Silverans}
{Silverans}, R.~E., {Borghs}, G., {de Bisschop}, P., \& {van Hove}, M. 1986, \pra, 33, 2117, \dodoi{10.1103/PhysRevA.33.2117}

\bibitem[{{Skrutskie} {et~al.}(2006){Skrutskie}, {Cutri}, {Stiening}, {Weinberg}, {Schneider}, {Carpenter}, {Beichman}, {Capps}, {Chester}, {Elias}, {Huchra}, {Liebert}, {Lonsdale}, {Monet}, {Price}, {Seitzer}, {Jarrett}, {Kirkpatrick}, {Gizis}, {Howard}, {Evans}, {Fowler}, {Fullmer}, {Hurt}, {Light}, {Kopan}, {Marsh}, {McCallon}, {Tam}, {Van Dyk}, \& {Wheelock}}]{2MASS}
{Skrutskie}, M.~F., {Cutri}, R.~M., {Stiening}, R., {et~al.} 2006, \aj, 131, 1163, \dodoi{10.1086/498708}

\bibitem[{{Smith} \& {Lambert}(1990)}]{SmithLambert90}
{Smith}, V.~V., \& {Lambert}, D.~L. 1990, \apjs, 72, 387, \dodoi{10.1086/191421}

\bibitem[{{Stancliffe}(2021)}]{Stancliffe21}
{Stancliffe}, R.~J. 2021, \mnras, 505, 5554, \dodoi{10.1093/mnras/stab1734}

\bibitem[{{Storm} \& {Bergemann}(2023)}]{2023Storm}
{Storm}, N., \& {Bergemann}, M. 2023, \mnras, 525, 3718, \dodoi{10.1093/mnras/stad2488}

\bibitem[{{Storm} {et~al.}(2024){Storm}, {Barklem}, {Yakovleva}, {Belyaev}, {Palmeri}, {Quinet}, {Lodders}, {Bergemann}, \& {Hoppe}}]{2024Storm}
{Storm}, N., {Barklem}, P.~S., {Yakovleva}, S.~A., {et~al.} 2024, \aap, 683, A200, \dodoi{10.1051/0004-6361/202348971}

\bibitem[{{Tout} \& {Eggleton}(1988)}]{ToutEggleton88}
{Tout}, C.~A., \& {Eggleton}, P.~P. 1988, MNRAS, 231, 823, \dodoi{10.1093/mnras/231.4.823}

\bibitem[{{Travaglio} {et~al.}(2004){Travaglio}, {Gallino}, {Arnone}, {Cowan}, {Jordan}, \& {Sneden}}]{Travaglio2004}
{Travaglio}, C., {Gallino}, R., {Arnone}, E., {et~al.} 2004, \apj, 601, 864, \dodoi{10.1086/380507}

\bibitem[{{Udry} {et~al.}(1998){Udry}, {Jorissen}, {Mayor}, \& {Van Eck}}]{Udry98}
{Udry}, S., {Jorissen}, A., {Mayor}, M., \& {Van Eck}, S. 1998, A\&AS, 131, 25, \dodoi{10.1051/aas:1998249}

\bibitem[{{Villemoes} {et~al.}(1992){Villemoes}, {Arnesen}, {Heijkenskj{\"o}ld}, {Kastberg}, \& {W{\"a}nnstr{\"o}m}}]{1992Villemoes}
{Villemoes}, P., {Arnesen}, A., {Heijkenskj{\"o}ld}, F., {Kastberg}, A., \& {W{\"a}nnstr{\"o}m}, A. 1992, Physics Letters A, 162, 178, \dodoi{10.1016/0375-9601(92)90998-2}

\bibitem[{{Vitali} {et~al.}(2024){Vitali}, {Slumstrup}, {Jofr{\'e}}, {Casamiquela}, {Korhonen}, {Blanco-Cuaresma}, {Winther}, \& {Aguirre B{\o}rsen-Koch}}]{2024Vitali}
{Vitali}, S., {Slumstrup}, D., {Jofr{\'e}}, P., {et~al.} 2024, \aap, 687, A164, \dodoi{10.1051/0004-6361/202349049}

\bibitem[{{Wendt} \& {Karstensen}(1984)}]{1984Wendt}
{Wendt}, H.~H., \& {Karstensen}, F. 1984, \pra, 29, 562, \dodoi{10.1103/PhysRevA.29.562}

\bibitem[{{Wright} {et~al.}(2010){Wright}, {Eisenhardt}, {Mainzer}, {Ressler}, {Cutri}, {Jarrett}, {Kirkpatrick}, {Padgett}, {McMillan}, {Skrutskie}, {Stanford}, {Cohen}, {Walker}, {Mather}, {Leisawitz}, {Gautier}, {McLean}, {Benford}, {Lonsdale}, {Blain}, {Mendez}, {Irace}, {Duval}, {Liu}, {Royer}, {Heinrichsen}, {Howard}, {Shannon}, {Kendall}, {Walsh}, {Larsen}, {Cardon}, {Schick}, {Schwalm}, {Abid}, {Fabinsky}, {Naes}, \& {Tsai}}]{WISE}
{Wright}, E.~L., {Eisenhardt}, P. R.~M., {Mainzer}, A.~K., {et~al.} 2010, \aj, 140, 1868, \dodoi{10.1088/0004-6256/140/6/1868}

\end{thebibliography}
\bibliographystyle{aasjournal}
\newpage

\begin{appendix}
\section{Line list}\label{sec:append}
\begin{table}[htbp]
    \centering
    \caption{Atomic lines used for deriving the abundance ratios presented in this work. Their logarithm values of the oscillator strength ($\mathrm{log}\,gf$) and excitation potentials ($E_{low}$) are sourced from the atomic database presented in the study by \cite{2021Heiter}}
    \label{tab:linelist}
    \scriptsize
    \begin{tabular}{|c|c|c|c|c|c|c|c|c|c|c|c|c|c|c|c|c|c|c|c|}
        \toprule
        Element & $\lambda$ (nm) & $E_{low}$& $\mathrm{log}\,gf$& Element & $\lambda$ (nm) & $E_{low}$& $\mathrm{log}\,gf$ &Element & $\lambda$ (nm) & $E_{low}$& $\mathrm{log}\,gf$ &Element & $\lambda$ (nm) & $E_{low}$& $\mathrm{log}\,gf$ &Element & $\lambda$ (nm) & $E_{low}$& $\mathrm{log}\,gf$  \\
        \midrule
        \begin{filecontents*}{line_list_sorted_new.csv}
1,2,3,4,5,6,7,8,9,a,b,c,d,e,f,g,h,i,l,m
Fe 1,491.801,4.231,-1.26,Fe 1,548.714,4.415,-1.43,Ni 1,568.22,4.105,-0.344,Ti 1,501.616,0.848,-0.48,Sc 2,531.835,1.357,-2.015
Fe 1,491.899,2.865,-0.342,Fe 1,549.183,4.186,-2.188,Ni 1,574.835,1.676,-3.24,Ti 1,521.97,0.021,-2.22,Sc 2,566.715,1.5,-1.309
Fe 1,494.564,4.209,-1.41,Fe 1,553.852,4.218,-1.54,Ni 1,585.775,4.167,-0.402,Ti 1,522.454,2.103,-0.28,Sc 2,566.904,1.5,-1.2
Fe 1,494.639,3.368,-1.11,Fe 1,553.928,3.642,-2.56,Ni 1,608.628,4.266,-0.41,Ti 1,542.914,2.345,-0.55,Sc 2,568.42,1.507,-1.074
Fe 1,495.76,2.808,0.233,Fe 1,554.995,3.695,-2.81,Ni 1,618.671,4.105,-0.88,Ti 1,550.39,2.578,-0.05,Y 2,488.368,1.084,0.19
Fe 1,496.257,4.178,-1.182,Fe 1,557.284,3.397,-0.289,Ni 1,620.46,4.088,-1.08,Ti 1,568.946,2.297,-0.36,Y 2,490.012,1.033,0.03
Fe 1,496.609,3.332,-0.792,Fe 1,557.31,4.191,-1.295,Ni 1,622.398,4.105,-0.91,Ti 1,592.211,1.046,-1.38,Y 2,508.742,1.084,-0.16
Fe 1,496.992,4.218,-0.71,Fe 1,558.676,3.368,-0.111,Ni 1,625.96,4.089,-1.237,Ti 1,597.854,1.873,-0.44,Y 2,520.041,0.992,-0.47
Fe 1,498.25,4.103,0.156,Fe 1,561.564,3.332,0.043,Ni 1,632.217,4.154,-1.115,Ti 1,609.117,2.267,-0.32,Y 2,532.078,1.084,-1.95
Fe 1,498.385,4.103,-0.006,Fe 1,561.863,4.209,-1.255,Ni 1,633.911,4.154,-0.53,Ti 1,612.622,1.067,-1.368,Y 2,572.889,1.839,-1.15
Fe 1,499.278,4.26,-2.35,Fe 1,562.454,3.417,-0.769,Ni 1,637.825,4.154,-0.82,Ti 2,541.877,1.582,-2.13,Mg 1,571.109,4.346,-1.724
Fe 1,500.404,4.209,-1.3,Fe 1,563.826,4.22,-0.72,Ni 1,641.458,4.154,-1.16,Cr 1,493.634,3.113,-0.25,Mg 1,631.872,5.108,-2.103
Fe 1,502.319,4.283,-1.5,Fe 1,566.134,4.284,-1.756,Ni 1,642.485,4.167,-1.355,Cr 1,495.372,3.122,-1.48,Mg 1,631.924,5.108,-2.324
Fe 1,502.813,3.573,-1.036,Fe 1,566.252,4.178,-0.447,Ca 1,526.17,2.521,-0.579,Cr 1,524.756,0.961,-1.59,Nd 2,508.983,0.205,-1.16
Fe 1,503.078,3.237,-2.83,Fe 1,570.546,4.301,-1.355,Ca 1,534.946,2.709,-0.31,Cr 1,527.2,3.449,-0.42,Nd 2,509.279,0.38,-0.61
Fe 1,503.191,4.371,-1.57,Fe 1,585.508,4.608,-1.478,Ca 1,551.298,2.933,-0.464,Cr 1,529.669,0.983,-1.36,Nd 2,527.687,0.859,-0.44
Fe 1,505.464,3.64,-1.921,Fe 1,593.465,3.929,-1.07,Ca 1,558.875,2.526,0.358,Cr 1,530.418,3.464,-0.67,Nd 2,529.316,0.823,0.1
Fe 1,507.922,2.198,-2.068,Fe 1,609.666,3.984,-1.83,Ca 1,559.011,2.521,-0.571,Cr 1,531.286,3.449,-0.55,Nd 2,531.981,0.55,-0.14
Fe 1,507.974,0.99,-3.221,Fe 1,618.02,2.728,-2.591,Ca 1,560.128,2.526,-0.523,Cr 1,531.877,3.438,-0.67,Ce 2,527.423,1.044,0.13
Fe 1,508.334,0.958,-2.939,Fe 1,618.799,3.943,-1.62,Ca 1,586.756,2.933,-1.57,Cr 1,532.978,2.914,-0.795,Ce 2,533.056,0.869,-0.4
Fe 1,509.077,4.256,-0.44,Fe 1,621.928,2.198,-2.432,Ca 1,610.272,1.879,-0.85,Cr 1,534.045,3.438,-0.73,Ce 2,547.228,1.247,-0.1
Fe 1,510.403,3.017,-2.77,Fe 1,624.065,2.223,-3.23,Ca 1,612.222,1.886,-0.38,Cr 1,534.476,3.449,-0.99,Ce 2,597.582,1.327,-0.45
Fe 1,512.462,3.301,-2.96,Fe 1,624.632,3.603,-0.771,Ca 1,615.602,2.521,-2.521,Cr 1,538.697,3.369,-0.743,Ce 2,604.337,1.206,-0.48
Fe 1,512.736,0.915,-3.306,Fe 1,625.256,2.404,-1.699,Ca 1,616.13,2.523,-1.266,Cr 1,562.864,3.422,-0.74,Ce 2,605.182,0.232,-1.53
Fe 1,522.24,2.279,-3.832,Fe 1,631.802,2.453,-1.804,Ca 1,616.217,1.899,-0.17,Cr 1,569.474,3.857,-0.241,Na 1,498.281,2.104,-0.916
Fe 1,524.249,3.634,-0.967,Fe 1,640.032,0.915,-4.318,Ca 1,616.376,2.521,-1.286,Cr 1,666.108,4.193,-0.113,Na 1,615.423,2.102,-1.547
Fe 1,524.378,4.256,-1.05,Fe 1,641.165,3.654,-0.596,Ca 1,616.644,2.521,-1.142,Si 1,564.561,4.93,-2.043,Na 1,616.075,2.104,-1.246
Fe 1,525.302,2.279,-3.84,Fe 1,641.995,4.733,-0.2,Ca 1,616.956,2.526,-0.478,Si 1,568.448,4.954,-1.553,Co 1,523.021,1.74,-1.84
Fe 1,525.346,3.283,-1.579,Fe 1,642.135,2.279,-2.012,Ca 1,643.908,2.526,0.39,Si 1,614.502,5.616,-1.31,Co 1,533.145,1.785,-2.782
Fe 1,526.727,4.371,-1.596,Fe 1,643.084,2.176,-2.005,Ca 1,644.981,2.521,-0.502,Si 1,623.732,5.614,-0.975,Co 1,564.723,2.28,-2.626
Fe 1,528.362,3.241,-0.452,Fe 1,660.802,2.279,-3.93,Ca 1,645.56,2.523,-1.29,Si 1,640.729,5.871,-1.393,Co 1,611.699,1.785,-2.695
Fe 1,529.878,3.642,-2.016,Fe 1,662.502,1.011,-5.336,Ca 1,647.166,2.526,-0.686,Si 1,672.185,5.863,-1.062,Al 1,669.602,3.143,-1.569
Fe 1,530.23,3.283,-0.738,Fe 1,667.798,2.692,-1.418,Ca 1,649.965,2.523,-0.818,Zr 1,481.504,0.651,-0.53,Al 1,669.867,3.143,-1.87
Fe 1,531.752,4.143,-2.462,Fe 1,671.032,1.485,-4.764,V 1,480.752,2.125,0.38,Zr 1,481.563,0.604,-0.03,Zn 1,481.053,4.078,-0.16
Fe 1,532.004,3.642,-2.44,Fe 1,675.271,4.638,-1.204,V 1,562.487,1.051,-1.06,Zr 1,482.804,0.623,-0.64,Pr 2,525.973,0.633,-0.539
Fe 1,532.418,3.211,-0.108,Fe 1,679.326,4.076,-2.326,V 1,562.763,1.081,-1.128,Zr 1,612.744,0.154,-1.06,Pr 2,532.277,0.483,-0.933
Fe 1,532.999,4.076,-1.196,Ni 1,494.544,3.796,-0.82,V 1,565.744,1.064,-1.4,Zr 1,613.455,0,-1.28,Cu 1,521.82,3.817,-0.095
Fe 1,533.993,3.266,-0.635,Ni 1,496.517,3.796,-1.212,V 1,566.836,1.081,-1.795,La 2,492.096,0.126,-0.58,Ba 2,585.367,0.604,-0.907
Fe 1,537.957,3.695,-1.514,Ni 1,497.613,3.606,-1.26,V 1,570.358,1.051,-0.851,La 2,512.299,0.321,-1.111,Eu 2,664.51,1.38,-0.625
Fe 1,538.633,4.154,-1.67,Ni 1,497.632,1.676,-3,V 1,572.765,1.051,-1.425,La 2,529.082,0,-1.65,,,,
Fe 1,539.317,3.241,-0.719,Ni 1,500.374,1.676,-3.07,V 1,573.706,1.064,-1.338,La 2,530.351,0.321,-1.731,,,,
Fe 1,541.278,4.435,-1.716,Ni 1,504.219,3.658,-0.58,V 1,603.972,1.064,-1.03,La 2,639.046,0.321,-2.012,,,,
Fe 1,542.407,4.32,0.52,Ni 1,509.993,3.679,-0.1,V 1,608.144,1.051,-0.582,Mn 1,482.352,2.319,-0.474,,,,
Fe 1,544.504,4.387,-0.02,Ni 1,511.539,3.834,-0.11,V 1,609.021,1.081,-0.827,Mn 1,537.761,3.844,-0.839,,,,
Fe 1,544.692,0.99,-1.914,Ni 1,522.029,3.74,-1.31,V 1,611.952,1.064,-0.7,Mn 1,541.367,3.859,-0.647,,,,
Fe 1,546.64,4.371,-0.63,Ni 1,539.233,4.154,-1.315,V 1,615.016,0.301,-2.264,Mn 1,543.254,0,-4.397,,,,
Fe 1,546.699,3.573,-2.233,Ni 1,546.249,3.847,-0.818,V 1,619.92,0.287,-2.323,Mn 1,547.063,2.164,-2.888,,,,
Fe 1,547.316,4.191,-2.04,Ni 1,558.786,1.935,-2.39,V 1,627.465,0.267,-1.967,Mn 1,601.349,3.072,-1.354,,,,
Fe 1,547.39,4.154,-0.72,Ni 1,561.477,4.154,-0.573,V 1,653.142,1.218,-1.22,Sc 2,503.102,1.357,-0.4,,,,
Fe 1,548.31,4.154,-1.392,Ni 1,566.994,4.266,-1.004,Ti 1,500.964,0.021,-2.2,Sc 2,523.981,1.455,-0.765,,,,
\end{filecontents*}
\csvreader[head to column names, late after line=\\]{line_list_sorted_new.csv}{}{\1 & \2 & \3 & \4 & \5 & \6 & \7 & \8 & \9 &\a  & \b & \c & \d & \e & \f & \g & \h & \i & \l & \m  }
        \bottomrule
    \end{tabular}
\end{table}

\end{appendix}

\end{document}